\documentclass[sigconf, nonacm]{acmart}
\usepackage{multirow}
\usepackage[ruled,vlined]{algorithm2e}
\usepackage{enumitem}
\usepackage{balance}
\usepackage{subfigure}
\usepackage{color}
\newtheorem{problem}{Problem}

\newcommand{\rev}[1]{\textcolor{black}{#1}}
\newcommand\vldbdoi{XX.XX/XXX.XX}
\newcommand\vldbpages{XXX-XXX}
\newcommand\vldbvolume{16}
\newcommand\vldbissue{1}
\newcommand\vldbyear{2022}
\newcommand\vldbauthors{\authors}
\newcommand\vldbtitle{\shorttitle} 
\newcommand\vldbavailabilityurl{URL_TO_YOUR_ARTIFACTS}
\newcommand\vldbpagestyle{plain} 

\begin{document}
\title{PromptEM: Prompt-tuning for Low-resource Generalized \\Entity Matching}
% \title{PromptEM: Prompt-tuning for Low-resource Generalized Entity Matching [Scalable Data Science]}

%%
%% The "author" command and its associated commands are used to define the authors and their affiliations.
\author{Pengfei Wang, Xiaocan Zeng, Lu Chen, Fan Ye, Yuren Mao, Junhao Zhu, Yunjun Gao}
\affiliation{%
  \institution{
    Zhejiang University, Hangzhou, China
    }
}

\email{{wangpf, zengxc, luchen, fan.ye, maoyuren, zhujunhao, gaoyj}@zju.edu.cn}

\begin{abstract}
% Entity Matching (EM) is one of the fundamental problems in data management, which aims to identify whether two entity records from two relational tables refer to the same real-world entity. Generalized Entity Matching (GEM) is a new research problem that generalizes traditional EM by allowing matching entity records of different formats. GEM can support more practical scenarios (e.g., paper matching). Existing supervised deep learning EM approaches rely on a large amount of high-quality labels examples to achieve considerable performance, however, the labeling process of GEM is extremely labor-intensive like EM. A low-resource but effective solution is an urgent need to advance real-world applications. To this end, we propose \textsf{PromptEM}, a novel low-resource GEM solution powered by prompt-tuning and self-training. Firstly, prompt-tuning is introduced to cast GEM as a cloze-style task, which bridges the gap between objectives forms in pre-training and fine-tuning. Thus we can stimulate the rich knowledge distributed in pre-trained language models (LMs) for better matching results. Then we develop a lightweight and efficient self-training method to further boost performance, which selects pseudo-labels by uncertainty along with pruning useless training data via the proposed MC-EL2N. Extensive experimental results on seven real-world benchmarks demonstrate the superiority of the proposed \textsf{PromptEM} against state-of-the-art approaches in terms of effectiveness and efficiency.

Entity Matching (EM), which aims to identify whether two entity records from two relational tables refer to the same real-world entity, is one of the fundamental problems in data management. Traditional EM assumes that two tables are homogeneous with the aligned schema, while it is common that entity records of different formats (e.g., relational, semi-structured, or textual types) involve in practical scenarios.
\rev{It is not practical to unify their schemas due to the different formats.} 
To support EM on format-different entity records, Generalized Entity Matching (GEM) has been proposed and gained much attention recently. To do GEM, existing methods typically perform in a supervised learning way, which relies on a large amount of high-quality labeled examples. However, the labeling process is extremely labor-intensive, and frustrates the use of GEM. Low-resource GEM, i.e., GEM that only requires a small number of labeled examples, becomes an urgent need. To this end, this paper, for the first time, focuses on the low-resource GEM and proposes a novel low-resource GEM method, termed as PromptEM. PromptEM has addressed three challenging issues (i.e., designing GEM-specific prompt-tuning, improving pseudo-labels quality, and running efficient self-training) in low-resource GEM.
Extensive experimental results on \rev{eight} real benchmarks demonstrate the superiority of PromptEM in terms of effectiveness and efficiency. 
\end{abstract}

\maketitle

\pagestyle{\vldbpagestyle}
\begingroup\small\noindent\raggedright\textbf{PVLDB Reference Format:}\\
\vldbauthors. \vldbtitle. PVLDB, \vldbvolume(\vldbissue): \vldbpages, \vldbyear.\\
\href{https://doi.org/\vldbdoi}{doi:\vldbdoi}
\endgroup
\begingroup
\renewcommand\thefootnote{}\footnote{\noindent
This work is licensed under the Creative Commons BY-NC-ND 4.0 International License. Visit \url{https://creativecommons.org/licenses/by-nc-nd/4.0/} to view a copy of this license. For any use beyond those covered by this license, obtain permission by emailing \href{mailto:info@vldb.org}{info@vldb.org}. Copyright is held by the owner/author(s). Publication rights licensed to the VLDB Endowment. \\
\raggedright Proceedings of the VLDB Endowment, Vol. \vldbvolume, No. \vldbissue\ %
ISSN 2150-8097. \\
\href{https://doi.org/\vldbdoi}{doi:\vldbdoi} \\
}\addtocounter{footnote}{-1}\endgroup

\ifdefempty{\vldbavailabilityurl}{}{
\vspace{.3cm}
\begingroup\small\noindent\raggedright\textbf{PVLDB Artifact Availability:}\\
The source code, data, and/or other artifacts have been made available at \url{https://github.com/ZJU-DAILY/PromptEM}.
\endgroup
}

\section{Introduction}
\label{sec:intro}
\begin{figure}
    \centering
    \includegraphics[width=3.3in]{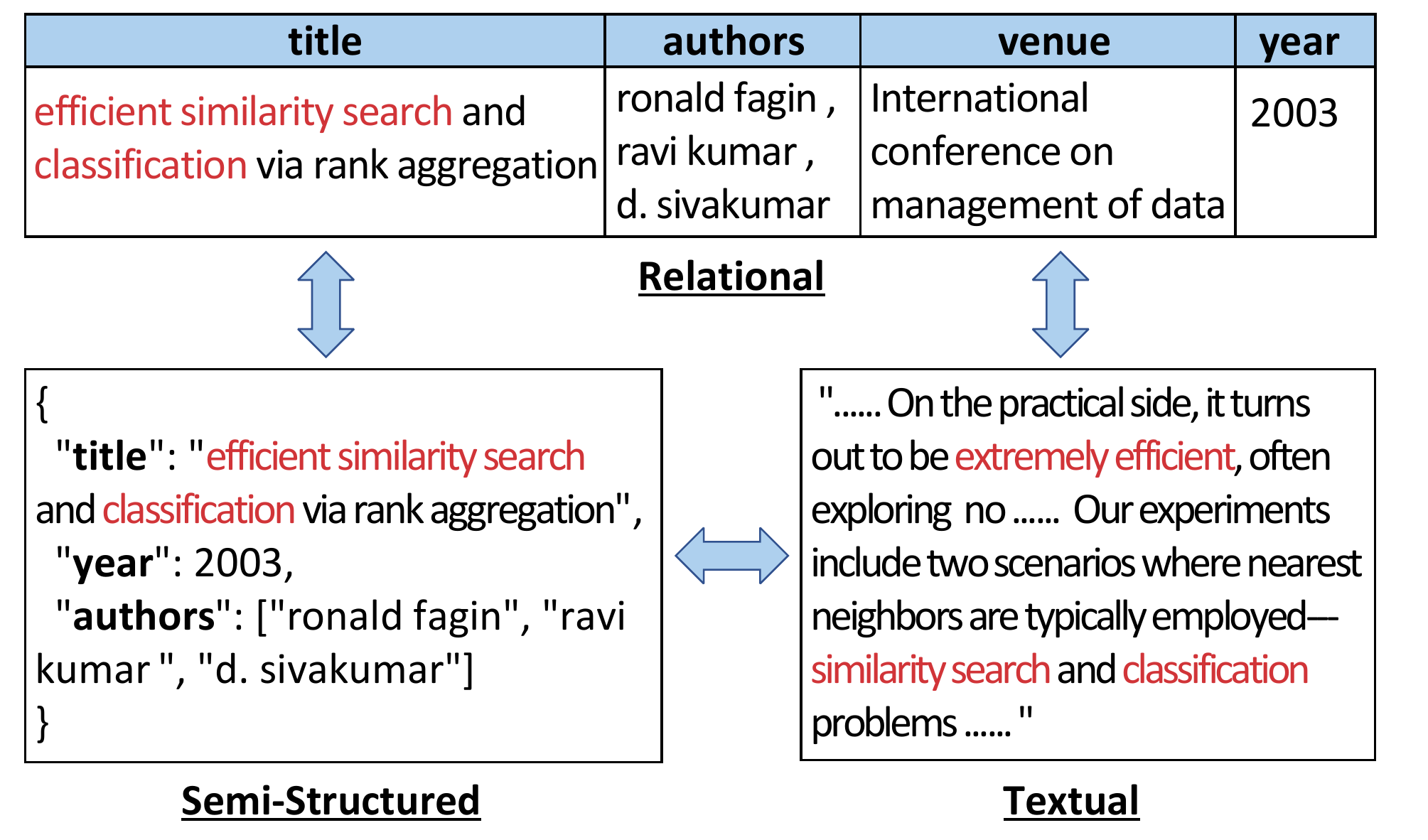}
    \vspace{-4.3mm}
    \caption{An example of generalized entity matching.}
    \label{fig:example}
    \vspace{-6mm}
\end{figure}

Entity Matching (EM) aims to identify whether two entity records from two relational tables refer to the same real-world entity, which is one of the fundamental and significant tasks in data management.
Most existing solutions \cite{mudgal2018deep, li2020deep, miao2021rotom} assume that two tables are homogeneous with the aligned schema.
However, there is an urgent need to generalize the entity matching problem for more practical scenarios. \rev{Taking paper matching as an example, as shown in Figure~\ref{fig:example}, paper metadata is usually stored in relational tables or semi-structured JSON objects, while paper description (e.g., abstract) is textual data. }
\rev{It is not practical to unify their schemas since we need a potentially expensive schema matching in the pre-processing step~\cite{madhavan2001generic}, which is even not applicable when matching data of different formats (e.g., matching paper metadata stored in tabular format with their textual abstracts).}
\rev{Therefore, traditional EM is unable to support those practical scenarios like paper matching~\cite{wang2021machamp}.}
\rev{Recently, \textsf{TDmatch} \cite{ahmadi2021unsupervised} first attempts to match structured data and textual data while having extremely poor scalability (to be analyzed via experiments in Section \ref{exp:efficiency}).}
To support more practical application scenarios, following the previous study \cite{wang2021machamp}, we study \emph{Generalized Entity Matching} (GEM), which allows matching entity records of relational, semi-structured, or textual types.

Existing methods designed for GEM typically perform in a supervised learning way, which relies on a large amount of labeled training data, and thus, is extremely labor-intensive. Recent studies~\cite{li2020deep, miao2021rotom} have achieved considerable performance by leveraging the power of pre-trained language models (LMs) and the fine-tuning paradigm. Nonetheless, the fine-tuning paradigm still requires a non-trivial amount of high-quality labeled examples (e.g., thousands of labels for a typical GEM application \cite{wang2021machamp}). 
\rev{\textsf{TDmatch}~\cite{ahmadi2021unsupervised} performs in an unsupervised learning way via graph creation and random walk. However, two drawbacks restrict its real-world applications as confirmed via experiments in Section \ref{sec:experiments}: (i) it has unstable performance due to the absence of labeled samples; and (ii) random walk is not scalable on large datasets~\cite{tong2006fast}, incurring enormous execution time and memory usage (e.g., more than 120 hours and 130 gigabytes for once training of the SEMI-REL).}
To satisfy real-life applications, a low-resource (i.e., using only a small number of labeled examples) but an effective and efficient solution is required. To the best of our knowledge, low-resource GEM remains unexplored.

To overcome the low-resource dilemma, semi-supervised learning techniques (e.g., self-training \cite{van2020uncertainty}) are good choice. Self-training has recently been shown to obtain state-of-the-art performance for low-resource tasks like sequence generation \cite{he2019revisiting} and speech recognition \cite{xu2021self}. In self-training, a \emph{teacher} model is trained on some labeled data, and is used to produce pseudo-labels on unlabeled data. Furthermore, the original labeled data is augmented with the pseudo-label data, and is employed to train a \emph{student} model. Thus, a large amount of unlabeled data can be utilized effectively and the requirement of labeled data is correspondingly reduced. Although self-training has achieved promising performance in a wide range of applications, it has not been explored in GEM.

% Self-training \cite{van2020uncertainty} as one of the promising semi-supervised learning approaches to address this shortcoming has recently been shown to obtain state-of-the-art performance for low-resource tasks like sequence generation \cite{he2019revisiting} and speech recognition \cite{xu2021self}. Formally, a \emph{teacher} model is trained on some labeled data and used to produce pseudo-labels on unlabeled data. The original labeled data is augmented with the pseudo-label data and used to train a \emph{student} model. So that lots of unlabeled data can be used effectively. 

\begin{figure*}
    \centering
    \includegraphics[width=7in]{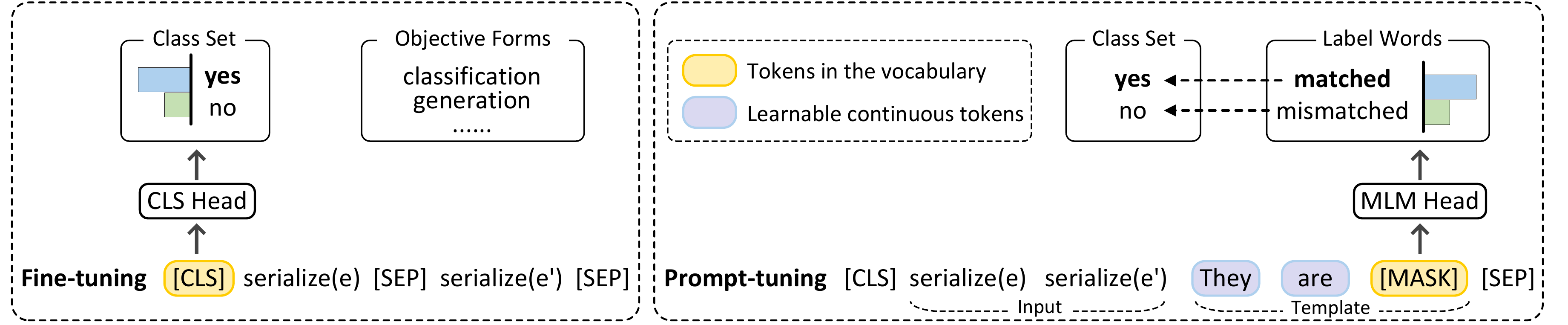}
    \vspace*{-7.5mm}
    \caption{The illustration of fine-tuning and prompt-tuning. The blue rectangles in the figure are special prompt tokens, whose parameters are initialized and learnable during prompt-tuning.}
    \label{fig:prompt}
    \vspace*{-5mm}
\end{figure*}

Motivated by the above considerations, we study the problem of learning high-quality models for low-resource GEM by means of self-training. Our goal is to develop an effective and efficient low-resource GEM solution based on pre-trained LMs along with leveraging self-training to boost performance, which is a challenging endeavor. The challenges are mainly three-folds:

\noindent\textbf{Challenge \uppercase\expandafter{\romannumeral 1}:} \emph{How to tune pre-trained LMs for GEM better?} Despite the success of fine-tuning LMs for the matching problem, some recent studies \cite{liu2021pre, liu2021gpt, han2021pre} find that there is a significant gap between objective forms in pre-training and fine-tuning, which restricts taking full advantage of knowledge in LMs. 
\rev{Pre-training is usually formalized as a cloze-style task to predict target words (e.g., masked language models). However, the approaches based on fine-tuning add additional layers to do different objective forms (e.g., classification and generation) as illustrated in Figure \ref{fig:prompt}. 
To do GEM, existing approaches treat it as a classification problem.
This gap hinders the transfer and adaptation of knowledge in LMs for GEM tasks.}

\noindent\textbf{Challenge \uppercase\expandafter{\romannumeral 2}:} \emph{How to select high-quality pseudo-labels?} The quality of pseudo-labels determines whether self-training can improve performance. Thus, the pseudo-label selection strategy is extremely important. A common strategy is using confidence to select pseudo-labels. However, this strategy has some serious drawbacks \cite{rizve2021defense, mukherjee2020uncertainty}. On the one hand, incorrect predictions can have high confidence scores in poorly calibrated networks. On the other hand, if we only aim at the pseudo-labels with high confidence produced by the teacher, there is little to gain for the student model.

\noindent\textbf{Challenge \uppercase\expandafter{\romannumeral 3}:} \emph{How to avoid expensive self-training?}
Traditional self-training can be costly. To be more specific, the labeled data is augmented by the pseudo-labels produced by the teacher model, which may be beneficial to performance but result in a longer training time. Intuitively, maybe not all training data contribute to boosting the performance of the student model. Nevertheless, how to quantify the importance of training data to avoid expensive self-training is still challenging.

To tackle the above three challenges, we propose a low-resource GEM solution \textsf{PromptEM}. \rev{Prompt-tuning is a new promising paradigm in natural language processing, and is able to bridge the gap of objective forms between pre-training and fine-tuning~\cite{liu2021pre, saunshi2020mathematical}.
To address the gap between pre-training and fine-tuning (\textbf{C\uppercase\expandafter{\romannumeral 1}}), we cast GEM as a cloze-style task via designing the GEM-specific prompt-tuning, which has the same objective form as pre-training. Thus, we can stimulate the rich knowledge distributed in LMs through prompt-tuning.} To select high-quality pseudo-labels (\textbf{C\uppercase\expandafter{\romannumeral 2}}), we develop a lightweight uncertainty-aware self-training method to boost performance. High-quality pseudo-labels are a prerequisite for boosting performance of self-training. To this end, we employ recent advances in Bayesian deep learning \cite{gal2016dropout} to obtain uncertainty estimates of the teacher model for pseudo-labeling and boosting the self-training process. To avoid expensive self-training (\textbf{C\uppercase\expandafter{\romannumeral 3}}), we prune useless training data dynamically using our proposed MC-EL2N, making the self-training process more lightweight and efficient. Our contributions are summarized as follows:
\begin{itemize}[topsep=0pt,itemsep=0pt,parsep=0pt,partopsep=0pt,leftmargin=*]
    \item \emph{Low-resource GEM.} This is the first work formally studying the problem of low-resource generalized entity matching. We articulate the importance of this problem in more practical scenarios.
    \item \rev{\emph{Prompt-tuning for GEM.} We present \textsf{PromptEM}, a new GEM solution based on prompt-tuning, which casts GEM as a cloze-style task. To the best of our knowledge, \textsf{PromptEM} is the first GEM (EM) solution that stimulates the rich knowledge distributed in LMs via designing GEM-specific prompt-tuning.}
    \item \emph{\rev{Generic Lightweight Self-training.}} \rev{To further improve the performance in low-resource settings, we develop a generic lightweight self-training method, which selects pseudo-labels using uncertainty and makes self-training more lightweight and efficient by dynamic data pruning.}
    \item \emph{Extensive Experiments.} \rev{We conduct comprehensive experimental evaluation on GEM tasks compared against state-of-the-art approaches, using eight real-world datasets from various domains.} Extensive experimental results demonstrate the superiority of our proposed \textsf{PromptEM} in terms of effectiveness and efficiency.
\end{itemize}

\noindent \textbf{Outline.} Section \ref{sec:preliminaries} presents the problem definition and overviews preliminaries. We introduce prompt-tuning for GEM in Section \ref{sec:prompt}. We further improve the performance by lightweight self-training in Section \ref{sec:self-training}. Section \ref{sec:experiments} presents the experimental results. Finally, we discuss related work in Section \ref{sec:related_work} and conclude in Section \ref{sec:conclusion}.

\vspace{-2.6mm}
\section{Preliminaries}
\label{sec:preliminaries}
In this section, we first present the problem definition of generalized entity matching (GEM). Next, we introduce the serializing method for GEM, followed by an introduction of conventional vanilla fine-tuning and prompt-based tuning with LMs.

\vspace{-2.5mm}
\subsection{Problem Formulation}
Given two collections of data entries,  entity matching (EM) is to identify pairs of data entries that refer to the same real-world entity.
A classic EM workflow \cite{konda2016magellan} has two main steps:
blocking and matching. The blocking \cite{thirumuruganathan2021deep} typically uses simple heuristics or deep learning techniques to reduce the quadratic number of candidates. The matching identifies whether each candidate pair is a real match or not. In this paper, we focus on the matching. Formally, given two tables $E_A$ and $E_B$, we assign a binary label $y \in \{0, 1\}$ for each candidate pair $(e_a, e_b) \in E_A \times E_B$. Here, $y=1$ denotes a truly matched pair, while $y=0$ represents a mismatched pair.

To generalize the classic setting to more practical application scenarios, \textsf{Machamp} \cite{wang2021machamp} comes up with the new research problem, generalized entity matching (GEM). 
GEM can support a variety of matching tasks with practical applications.
% GEM can support a variety of matching tasks with practical applications by allowing matching entity records of relational, semi-structural, or textual types.
\vspace{-2mm}
\begin{problem}
Generalized Entity Matching (GEM). Given two structured, semi-structured, or unstructured entity tables $E_A$ and $E_B$ with homogeneous or heterogeneous schema, GEM is to assign a binary label $y \in \{0, 1\}$ for each candidate $(e_a, e_b)
\in E_A \times E_B$.
\end{problem}
\vspace{-2mm}
\subsection{Serializing}
\label{sec:serializing}
The matching problem can be effectively solved by formulating it as a sequence classification task \cite{li2020deep, ge2021collaborem, wang2021machamp}. First, entity pairs are serialized to sequences, and then, a pre-trained LM is fine-tuned to solve the task. Existing methods are designed for EM over structured data with homogeneous data, which are not suitable for GEM. Following \cite{wang2021machamp}, we extend the serialization method presented in \textsf{Ditto} \cite{li2020deep} and introduce a reasonable way to fulfill this task.

\noindent \textbf{Structured tables.}
For structured tables, an entity with $n$ attributes can be denoted as $e=\left\{\operatorname{attr}_{i}, \mathrm{val}_{i}\right\}_{i \in[1, n]}$, where $\operatorname{attr}_i$ is the attribute name and $\operatorname{val}_i$ is the corresponding attribute value. Then the serialization is denoted as:

\noindent $serialize(e)::= \operatorname{[COL] \, attr_1 \, [VAL] \, val_1 \, ... \, [COL] \, attr_n \, [VAL] \, val_n}$

\noindent where $\operatorname{[COL]}$ and $\operatorname{[VAL]}$ are two special tags indicating the start of attribute names and values respectively. Taking the relational entity in Figure \ref{fig:example} as an example, we serialize it as:

\noindent $\operatorname{[COL] \, title \, [VAL] \, efficient \, similarity \, ... \, [COL] \, authors \, [VAL]}$

\noindent $\operatorname{renald \, ... \, [COL] \, venue \, [VAL] \, SIGMOD \, [COL] \, year \, [VAL] \, 2003}$

\noindent \textbf{Semi-structured tables.}
The semi-structured tables can be serialized in a similar way. Specially, two differences exist:
(i) For nested attributes, we recursively add the $\operatorname{[COL]}$ and $\operatorname{[VAL]}$ tags along with attribute names and values in each level of nests. (ii) \rev{To reduce the length of the sequence while ensuring the amount of information, we concatenate the elements in the list into one string for attributes whose content is a list.}
As an example, given the semi-structured entity in Figure \ref{fig:example}, we serialize it as:

\noindent $\operatorname{[COL] \, title \, [VAL] \, efficient \, similarity \, ... \, [COL] \, year \, [VAL] \, 2003}$

\noindent $\operatorname{[COL] \, authors \, [VAL] \, ronald \, fagin \, ravi \, kumar \, d.sivakumar}$

\noindent \textbf{Unstructured tables.}
Unstructured textual entities are sequences originally, and hence, there does not need to serialize them.
\vspace{-2.5mm}
\subsection{Vanilla Fine-tuning}
\rev{Pre-trained language models (LMs) (e.g., BERT \cite{devlin2018bert} and RoBERTa~\cite{liu2019roberta}) have demonstrated powerful semantic expression abilities, which can support lots of downstream tasks (e.g., classification and question answering).}
Formally, GEM can be treated as a sequence pair classification task denoted as $\mathcal{T}=\{\mathcal{X}, \mathcal{Y}\}$, where $\mathcal{X}$ is a candidate pair set and $\mathcal{Y}$ is a class set. For each instance $x \in \mathcal{X}$, it is serialized by $x::=[\mathrm{CLS}] \operatorname{serialize}(e) [\mathrm{SEP}] \operatorname{serialize}(e^{\prime}) [\mathrm{SEP}]$, and is annotated with a label $y \in \mathcal{Y}$.
Here, $\operatorname{[CLS]}$ and $\operatorname{[SEP]}$ are special tokens used to mark the beginning and end of a sequence.
% \footnote{Note that [CLS] and [SEP] are the special tokens for BERT and other LMs may have other special tokens, which are usually implemented as part of their tokenizers.}

Given a LM $\mathcal{M}$, vanilla fine-tuning first coverts $x$ to the input sequence $\left\{[\mathrm{CLS}], w_{1}, \ldots, w_{n},[\mathrm{SEP}], {w^{\prime}}_{1}, \ldots, {w^{\prime}}_{m}, [\mathrm{SEP}]\right\}$, and then, it uses $\mathcal{M}$ to encode all tokens of the input sequence into corresponding vectors $\left\{\mathbf{h}_{[\mathrm{CLS}]}, \mathbf{h}_{w_{1}}, \ldots, \mathbf{h}_{w_{n}}, \mathbf{h}_{[\mathrm{SEP}]}, \mathbf{h}_{{w}^{\prime}_{1}}, \ldots, \mathbf{h}_{{w}^{\prime}_{m}},
\mathbf{h}_{[\mathrm{SEP}]}\right\}$, where $w_{i}$ is the token and $\mathbf{h}_{w_{i}}$ is the corresponding embedding.
For a downstream classification task (e.g., GEM), a
task-specific head is trained to predict the probability distribution over the label set $y$ with the softmax function $p(\cdot \mid x)=\operatorname{Softmax}\left(\mathbf{W} \times \mathbf{h}_{[\text {CLS}]}+\mathbf{b}\right)$. Here, $\mathbf{h}_{[\text {CLS}]}$ is the embedding of special classification token \text{[CLS]}, $b$ is the bias for the layer, and $\mathbf{W}$ is a learnable matrix randomly initialized before fine-tuning. The parameters of $\mathcal{M}$, $b$, and $\mathbf{W}$ are tuned to maximize $\frac{1}{|\mathcal{X}|} \sum_{x \in \mathcal{X}} \log p\left(y \mid x\right)$.
\vspace{-2.5mm}
\subsection{Prompt-based Tuning}

Prompt-based tuning has been proposed to apply cloze-style tasks to tune LMs. Formally, we define a label word set $\mathcal{V}_{y}=\left\{w_{1}, \ldots, w_{m}\right\}$. $\mathcal{V}_{y}$ is a subset of the vocabulary $\mathcal{V}$ of the LM $\mathcal{M}$, i.e., $\mathcal{V}_{y} \subseteq \mathcal{V}$.
We get an overall dictionary $\mathcal{V}^{*}$ by taking the union of the dictionary corresponding to each label. Another primary component of prompt-tuning is a prompt template $T(\cdot)$, which modifies the original input $x$ into a prompt input $T(x)$ by adding a set of additional tokens in $x$.
Generally, a token $[\operatorname{MASK}]$ is added for LMs to predict the missing label word $w \in \mathcal{V}^{*}$. Thus, in prompt-tuning, a classification problem is transferred into a masked language modeling problem
$p(y \in \mathcal{Y} \mid x)=p\left([\operatorname{MASK}]=w \in \mathcal{V}_{y} \mid T(x)\right)$.

\vspace{-2mm}
\section{Prompt Tuning for GEM}
\label{sec:prompt}
\rev{In this section, we detail how to utilize prompt-tuning to deal with GEM. We first design GEM-specific prompt templates and label words, and then, we describe the training and inference process.}
\vspace{-2.5mm}
\subsection{Prompt Templates} 
% The goal of generalized entity matching is to predict if two heterogeneous or homogeneous entities are matching, and the answers can be "yes" or "no". Thus, we can construct templates based on the task definition. 
\rev{%Different tasks require different prompt templates and label words.
To cast the GEM problem as a prompt-tuning one, we first design suitable prompt templates (i.e., hard-encoding templates and continuous templates) and label words set (to consider general binary relationship).}

\noindent \textbf{Hard-encoding templates.} For the choice of hard-encoding templates, we do not use automatic searching methods for discrete prompts since the GEM task is clearly defined and the prompts are easily purposeful. Given each candidate pair $x=(e, e^\prime)$, we construct the following templates:

$\mathrm{T}_{1}(x)=serialize(e) \ \ serialize(e^\prime) \ \ \operatorname{They \ \ are \ \ [MASK]}  $

$\mathrm{T}_{2}(x)=serialize(e) \operatorname{\ \ is \ \ [MASK] \ \ to \ \ }  serialize(e^\prime)  $

\noindent \textbf{Continuous templates.} As prompt construction is to find a method that allows a LM to effectively perform a task, rather than being for human consumption, it is not necessary to limit the prompt to human-interpretable natural language~\cite{liu2021pre}. Thus, continuous prompts are proposed to perform prompting directly in the embedding space of the model. Here, we employ P-tuning~\cite{liu2021gpt}, where continuous prompt tokens are learned by inserting trainable variables into the embedding input. Specifically, trainable prompt tokens are initialized, and then, BiLSTM \cite{graves2013speech} is utilized to account for interaction between prompt tokens. This enables the model to find better continuous prompts beyond the original vocabulary $\mathcal{V}$ of $\mathcal{M}$ could express. We give an illustrative example in Figure \ref{fig:prompt}.

\noindent \textbf{Label words set.}
\rev{In addition to designing templates, another primary component is to design the label words set.
Note that, traditional EM tasks find pairs of entities that are identical \cite{wang2021machamp}.
However, GEM might require finding out entity pairs satisfying a general binary relationship.
Taking paper matching as an example, our goal is to find pairs between paper metadata and abstracts. Indeed, the relationship between them is whether they are relevant, which is more general beyond matching. Considering general binary relationship, we map the label $y=\text{yes}$ into a set $\mathcal{V}_{y}=\{\textit{matched}, \textit{similar}, \textit{relevant}\}$. Similarly, the label set for label $y=\text{no}$ is $\mathcal{V}_{y}=\{\textit{mismatched}, \textit{different}, \textit{irrelevant}\}$.}

\vspace{-2mm}
\subsection{Training and Inference}
A classification problem is transferred into a masked language modeling problem via prompt-tuning. In masked language modeling, we use confidence scores of all the words in $\mathcal{V}_{y}$ to construct the final score of the particular class $y$. Given a candidate pair $x$ (which is mapped to $T(x)$) and its class $y$ (which is mapped to $\mathcal{V}_{y}=\left\{w_{1}, \ldots, w_{m}\right\}$), the conditional probability is computed as:
\begin{equation}
\label{equ_1}
\setlength{\abovedisplayskip}{1pt}
\setlength{\belowdisplayskip}{1pt}
P(y \mid x)=\frac{1}{m} \sum_{j}^{m} P\left([\mathrm{MASK}]=w_{j} \mid T(x)\right)
\end{equation}

\noindent \textbf{Training.} The continuous prompt tokens can be parameterized by $\phi$ and optimized along with $\mathcal{M}$ during training. We tune the pre-trained model $\mathcal{M}$ (parameterized by $\theta$) along with the additional prompt embeddings by using the cross-entropy loss function $\mathcal{L}=-\sum \log P(y \mid x ; \theta, \phi)$. Here, we prompt-tune a pre-trained LM for the GEM task as follows:

\begin{enumerate}[topsep=0pt,itemsep=0pt,parsep=0pt,partopsep=0pt,leftmargin=*]
    \item Design task-specific prompt templates and label words set.
    \item Initialize the network with parameters from the pre-trained LM and continuous prompt tokens.
    \item Train the network on the training set until convergence.
\end{enumerate}
% (i) Design task-specific prompt templates and label words set.

% (ii) Initialize the network with parameters from the pre-trained LM and continuous prompt tokens.

% (iii) Train the network on the training set until convergence.

\rev{Different tasks require different prompt templates and label words. 
Step (1) is specifically designed for GEM tasks.
Continuous prompt tokens in Step (2) are specifically designed to enable the model to find better prompts beyond $\mathcal{V}$ of $\mathcal{M}$ could express.}

\noindent \textbf{Inference.} For inference, we aim to assign a label for the input, which can directly use Eq. \ref{equ_1} to predict the class of the current input instance based on predicted words of the $\operatorname{[MASK]}$ position. 
\vspace{-2mm}
\section{Lightweight Self-Training}
\label{sec:self-training}
With prompt-tuning, we can stimulate the rich knowledge distributed in LMs, which achieves considerable performance under low-resource settings. 
\rev{To further improve the performance and avoid expensive self-training, we develop a generic lightweight self-training method.}
%We first overview the lightweight self-training in this section, and then, we present two key modules: uncertainty-aware pseudo-label selection and dynamic data pruning.
\vspace{-2mm}
\subsection{Overview}
Let $D_{L}=\left\{\left(x^{(i)}, y^{(i)}\right)\right\}_{i=1}^{N_{L}}$ and $D_{U}=\left\{x^{(i)}\right\}_{i=1}^{N_{U}}$ be a labeled dataset with $N_L$ samples and an unlabeled dataset with $N_U$ samples, respectively.
\rev{Our lightweight self-training aims to boost the performance (i.e., effectiveness) using uncertainty meanwhile being more efficient and lightweight than traditional self-training via dynamic data pruning.}
We describe the lightweight self-training procedure, with its pseudo code presented in Algorithm \ref{algorithmn}. Given a labeled dataset $D_L$, a teacher model $\mathcal{M}_t$ is initialized and trained on $D_L$ until convergence (Lines 2-4). Then the teacher model $\mathcal{M}_t$ produces pseudo-labels on $D_U$. After that, we introduce an uncertainty-aware pseudo-label selection strategy to select high-quality pseudo-labels $D_P$ (Lines 5-6). Meanwhile, $D_U$ and $D_L$ are updated (Lines 7-8). Next, a student model $\mathcal{M}_s$ is initialized and trained on the updated $D_L$ (Lines 9-11). To make self-training more lightweight and efficient, we present a dynamic data pruning strategy, which can prune useless samples and their labels in $D_L$ every fixed number of epochs (Lines 12-15). Finally, we choose the best student model with the best performance on the validation set (Line 16). 
\rev{Since LST is general enough to incorporate with other approaches, it is possible to be widely used in practical low-resource applications.}
%Noted that, the above steps can repeat iteratively.
\SetAlgoSkip{}
\begin{algorithm}[t]
\caption{Lightweight Self-training}
\LinesNumbered
\DontPrintSemicolon
\label{algorithmn}
    \KwIn{the number $\mathbf{Iter}$ of iterations, a labeled train set $D_L$, an unlabeled train set $D_U$}
    \KwOut{a student model $\mathcal{M}_{\theta,\phi}$}
    \For{$i\gets 1$ \ \KwTo \ $\mathbf{Iter}$}{
        Initialize a new teacher model $\mathcal{M}_{t,\theta,\phi,i}$\;
        \For{$epoch\gets 1$ \ \KwTo \ Epochs of teacher}{
            Train $\mathcal{M}_{t,\theta,\phi,i}$ using the train set $D_L$\;
        }
        $\triangleright$ Uncertainty-aware Pseudo-label Selection \;
        $D_P \leftarrow$ Select pseudo-labels from $D_U$ \;
        $D_U \leftarrow D_U \ - \  D_P$ \;
        $D_L \leftarrow D_L \ \cup \  D_P$ \;
        Initialize a new student model $\mathcal{M}_{s,\theta,\phi,i}$\;
        \For{$epoch\gets 1$ \ \KwTo \ Epochs of student}{
            Train $\mathcal{M}_{s,\theta,\phi,i}$ using the train set $D_L$\;
            $\triangleright$ Dynamic Data Pruning \;
            \If{$( epoch \mod $ frequency of pruning $)=0$}{
                $D_D \leftarrow$ Select useless samples from $D_L$ \;
                $D_L \leftarrow D_L \ - \  D_D$ \;
            }
        }
    }
%$\triangleright$ return the best student model \;
\Return{the best student model $\mathcal{M}_{\theta,\phi}$}
\end{algorithm}

\vspace{-2.5mm}
\subsection{Uncertainty-aware Pseudo-label Selection}
Selecting high-quality pseudo-labels is a prerequisite for boosting self-training performance. Therefore, we aim at reducing the noise present in the selected samples to improve the overall performance. A straightforward way to select pseudo-labels is by selecting samples with high-confidence predictions. However, incorrect predictions can have high confidence scores in poorly calibrated networks~\cite{rizve2021defense}. Besides, if the teacher model already predicts some samples with high confidence, then these is little to gain for student model with these samples~\cite{mukherjee2020uncertainty}. Based on the observation that prediction uncertainties can be leveraged to negate the effect of poor calibration \cite{rizve2021defense}, we employ an uncertainty-aware pseudo-label selection strategy. Formally, uncertainty can be divided into \emph{epistemic uncertainty} and \emph{aleatoric uncertainty} \cite{van2020uncertainty}. The former comes from uncertainty in the parameters of the model, and the latter is uncertainty inherent in the data (e.g., two samples of different classes are similar). We focus on quantifying \emph{epistemic uncertainty}. Inspired by \cite{rizve2021defense, mukherjee2020uncertainty}, we use MC-Dropout \cite{gal2016dropout} to obtain an uncertainty measure by calculating the standard deviation of a fixed number (e.g., 10 in our experiments) of stochastic forward passes. To avoid carefully chosen thresholds, we choose $N_P$ samples with the least uncertainty after calculating the uncertainties of $D_U$:
\begin{equation}
    \setlength{\abovedisplayskip}{1pt}
    \setlength{\belowdisplayskip}{1pt}
    D_{P}=\left\{\left(x^{(i)}, \tilde{y}^{(i)}\right)\right\}_{i=1}^{N_{P}}=\operatorname{Top\,-}N_P\,(D_U\,|\,-u^{(i)})
\end{equation}
Here, $N_P=N_U \cdot u_r$, $u^{(i)}$ is the uncertainty of the sample, $\tilde{y}^{(i)}$ is the pseudo-label produced by the teacher model, and $u_r$ is the proportion of the unlabeled samples. The time complexity of the uncertainty estimation is $O(|N_P| \times log(|N_U|))$.
\rev{Uncertainty-aware pseudo-label selection makes the self-training process more effective, which will be analyzed in Section \ref{sec:variants}.}
\vspace{-2mm}
\subsection{Dynamic Data Pruning}
Traditional self-training can be expensive as the training set grows, resulting in more training time. Recently, Paul et al. \cite{paul2021deep} show that the Error L2-Norm (EL2N) score can identify important examples early in training. In other words, it can prune significant fractions of useless training data without sacrificing test accuracy, which can reduce the number of the training set and training time. Inspired by EL2N \cite{paul2021deep} and MC-Dropout \cite{gal2016dropout}, we combine these approaches and propose MC-EL2N, which is able to quantify importance scores more stably. Formally, the MC-EL2N score of a training sample $(x, y)$ is defined as ${\frac{\sum_{i=1}^n||\mathcal{M}(x)-y||_{2}}{n}}$, where $n$ is the number of stochastic forward passes. Similarly, to avoid carefully chosen thresholds, we choose $N_D$ samples with the least MC-EL2N score after quantifying the importance of $D_L$:
\begin{equation}
    \setlength{\abovedisplayskip}{1pt}
    \setlength{\belowdisplayskip}{1pt}
    D_{D}=\left\{\left(x^{(i)}, y^{(i)}\right)\right\}_{i=1}^{N_{D}}=\operatorname{Top\,-}N_D\,(D_L\,|\,-e^{(i)})
\end{equation}
Here, $N_D=N_L \cdot e_r$, $e^{(i)}$ is the MC-EL2N score of the sample, and $e_r$ is the proportion of the labeled samples. Similar to the uncertainty estimation, this process can be computed efficiently in $O(|N_D| \times log(|N_L|))$ time. 
\rev{We prune the useless samples every fixed epochs using dynamic data pruning, making the self-training process more lightweight and efficient. We will confirm the efficiency of dynamic data pruning, to be presented in Section \ref{exp:efficiency}.}

% \begin{table}[]
% \centering
% \caption{\rev{Statistics of the datasets used in our experiments.}}
% \vspace{-4mm}
% \label{table:dataset}
% \setlength{\tabcolsep}{0.6mm}{
% \begin{tabular}{c|cccc|ccc}
% \hline
% \multirow{2}{*}{Dataset} & \multicolumn{2}{c}{Left Table} & \multicolumn{2}{c|}{Right Table} & \multicolumn{3}{c}{Labeled Ground Truth} \\
%             & \#row  & \#attr & \#row  & \#attr & Num  & \% rate & Train \\ \hline
% REL-HETER   & 534    & 6.00   & 332    & 7.00   & 567     & 10\% & 57   \\
% SEMI-HOMO   & 2,616  & 8.65   & 64,263 & 7.34   & 17,223  & 5\%   & \\
% SEMI-HETER   & 22,133  & 12.28   & 23,264 & 12.03   & 1,240   & 10\% &    \\
% SEMI-REL    & 29,180 & 8.00   & 32,823 & 13.81  & 1,309   & 10\%  &  \\
% SEMI-TEXT-w & 9,234  & 10.00  & 9,234  & 1.00   & 5,540   & 10\%  & \\
% SEMI-TEXT-c & 20,897 & 10.00  & 20,897 & 1.00   & 12,538   & 5\%  &  \\
% REL-TEXT    & 2,616  & 1.00   & 2,295  & 6.00   & 7,417    & 10\%  & \\ \hline
% \end{tabular}}
% \vspace*{-5mm}
% \end{table}

\begin{table}[] \small
\centering
\caption{\rev{Statistics of the datasets used in our experiments. "All" denotes the total number of labeled samples, and "Train" represents the number of training samples used in our default low-resource setting.}}
\vspace{-4mm}
\label{table:dataset}
\setlength{\tabcolsep}{0.5mm}{
\begin{tabular}{|c|c|cccc|ccc|}
\hline
\multirow{2}{*}{Datasets} & \multirow{2}{*}{Domain} & \multicolumn{2}{c}{Left Table} & \multicolumn{2}{c|}{Right Table} & \multicolumn{3}{c|}{Labeled Examples} \\
                         &                         & \#row          & \#attr        & \#row           & \#attr         & All          & \% rate      & Train      \\ \hline
REL-HETER                &  restaurant   & 534            & 6.00          & 332             & 7.00           & 567          & 10\%         & 57         \\
SEMI-HOMO       & citation                         & 2,616          & 8.65          & 64,263          & 7.34           & 17,223       & 5\%          & 861           \\
SEMI-HETER               &  book                       & 22,133         & 12.28         & 23,264          & 12.03          & 1,240        & 10\%         & 124           \\
SEMI-REL                 &  movie                        & 29,180         & 8.00          & 32,823          & 13.81          & 1,309        & 10\%         & 131           \\
SEMI-TEXT-w              &  product                       & 9,234          & 10.00         & 9,234           & 1.00           & 5,540        & 10\%         & 554           \\
SEMI-TEXT-c              &  product                       & 20,897         & 10.00         & 20,897          & 1.00           & 12,538       & 5\%          & 627           \\
REL-TEXT                 &  citation                    & 2,616          & 1.00          & 2,295           & 6.00           & 7,417        & 10\%         &  742          \\
GEO-HETER                &  geo-spatial                    & 2,469          & 5.00          & 2,788           & 4.00           & 2,500        & 10\%         &  250          \\\hline
\end{tabular}}
\vspace{-6mm}
\end{table}
\vspace{-2mm}
\section{Experiments}
\label{sec:experiments}
In this section, we experimentally evaluate the proposed \textsf{PromptEM} on \rev{eight} real-world datasets. We aim at answering the following research questions:
\begin{itemize}[topsep=0pt,itemsep=0pt,parsep=0pt,partopsep=0pt,leftmargin=*]
    \item \textbf{RQ1}: \rev{How does \textsf{PromptEM} perform compared with the state-of-the-art methods under low-resource settings?}
    \item \textbf{RQ2}: How does each module affect the overall performance of the model?
    \item \textbf{RQ3}: How does \textsf{PromptEM} perform compared with state-of-the-art approaches in terms of efficiency?
    \item \textbf{RQ4}: Why do we choose these key modules (i.e., continuous templates and uncertainty-aware pseudo-label selection)?
\end{itemize}
\vspace{-2.5mm}
\subsection{Experimental Setup}

\noindent \textbf{Dataset.}
\rev{We use all the seven real-world benchmark datasets with different structures from \textsf{Machamp} \cite{wang2021machamp} and one geospatial dataset (GEO-HETER)~\cite{balsebre2022geospatial}. The detailed GEO-HETER construction can be found in our online version\footnote{\url{https://arxiv.org/pdf/2207.04802.pdf}}.}
The statistics of datasets are summarized in Table \ref{table:dataset}. 
\rev{Each dataset consists of the left and right tables of entities with possibly different formats (i.e., relational (REL) format, semi-structured (SEMI) format, or textual (TEXT) format).}
When they are of the same format, they can have a \rev{homogeneous} (HOMO) or heterogeneous (HETER) schema. 
%\rev{Following \cite{wang2021machamp}, we construct GEO-HETER using the dataset OSM-FSQ-Pittsburgh from \cite{balsebre2022geospatial}.}
We use rate\% of labeled data as training set (e.g., 57 labeled data for REL-HETER), and use the same train/valid/test splits as \textsf{Machamp}.

\noindent \textbf{Baselines.} 
We compare \textsf{PromptEM} with \rev{eight} SOTA EM methods, among which three (i.e., \textsf{Ditto}, \textsf{DADER}, and \textsf{Rotom}) have made efforts to \emph{low-resource EM}, and \textsf{TDmatch} is an unsupervised matching method for structural and textual data. 
%We summarize these methods below:

\begin{itemize}[topsep=0pt,itemsep=0pt,parsep=0pt,partopsep=0pt,leftmargin=*]
    \item \textbf{DeepMatcher} \cite{mudgal2018deep} is an entity matching framework that uses RNN architecture to aggregate the attribute values and then align the aggregated representations of the attributes.
    \item \textbf{BERT} \cite{devlin2018bert} is fine-tuned to treat GEM as a sequence pair classification task.
    \item \textbf{SentenceBERT} \cite{reimers2019sentence} proposes a siamese architecture for pre-trained LMs for sentence matching, and could also be applied to the task of GEM.
    \item \textbf{Ditto} \cite{li2020deep} is the SOTA EM approach that fine-tunes a pre-trained LM with three optimizations (i.e., domain knowledge, TF-IDF summarization, and data augmentation).
    \item \textbf{DADER} \cite{tu2022domain} presents a transfer learning based EM framework via domain adaptation.
    \item \textbf{Rotom} \cite{miao2021rotom} proposes a meta-learning framework that selects and weights the augmented data to better fine-tune the LMs.
    \item \textbf{TDmatch} \cite{ahmadi2021unsupervised} is an unsupervised approach to match textual and structured data using graph creation and random walk. \rev{Furthermore, we build an MLP classifier on top of its embeddings to perform in the supervised setting, called \textbf{TDmatch*}.}
\end{itemize}
% (i) \emph{DeepMatcher} ; (ii) \emph{BERT}; (iii) \emph{SentenceBERT} ; (iv) \emph{Ditto}; (v) \emph{HIF-KAT}; (vi) \emph{Rotom}; (vii) \emph{TDmatch} .

% We also compare with variants of \textsf{PromptEM} without the prompt-tuning (PT) and/or self-training (ST) to evaluate the effectiveness of each component.

\begin{table*}[h] \small
\centering
\caption{\rev{Results of all the methods under the default low-resource setting.}}
\vspace{-4.5mm}
\label{table:low}
\setlength{\tabcolsep}{0.8mm}{
\begin{tabular}{|c|ccc|ccc|ccc|ccc|ccc|ccc|ccc|ccc|}
\hline
\multirow{2}{*}{\textbf{Methods}} &
  \multicolumn{3}{c|}{\textbf{REL-HETER}} &
  \multicolumn{3}{c|}{\textbf{SEMI-HOMO}} &
  \multicolumn{3}{c|}{\textbf{SEMI-HETER}} &
  \multicolumn{3}{c|}{\textbf{SEMI-REL}} &
  \multicolumn{3}{c|}{\textbf{SEMI-TEXT-c}} &
  \multicolumn{3}{c|}{\textbf{SEMI-TEXT-w}} &
  \multicolumn{3}{c|}{\textbf{REL-TEXT}} &
  \multicolumn{3}{c|}{\textbf{GEO-HETER}}\\
                 & P     & R     & F     & P    & R    & F    & P    & R    & F    & P    & R    & F    & P    & R    & F    & P    & R     & F    & P    & R    & F  & P    & R    & F  \\ \hline
\textsf{DeepMatcher}      & 0.0   & 0.0   & 0.0   & 74.6 & 72.9 & 73.8 & 78.3 & 22.6 & 35.1 & 70.1 & 44.8 & 54.7 & 23.0 & 39.6 & 29.1 & 23.5 & 1.9   & 3.5  & 36.9 & 17.1 & 23.4 & 28.9 & 90.9 & 43.8 \\
\textsf{BERT}      & 100   & 90.9  & 95.2  & 90.1 & 93.2 & 91.6 & 63.6 & 17.6 & 27.6 & 92   & 94.5 & 93.3 & 56.9 & 47.2 & 51.6 & 19.6 & 20.9  & 20.2 & 27.1 & 26.6 & 26.9 & 70.8 & 90.4 & 79.4 \\
\textsf{SentenceBERT}     & 100   & 90.9  & 95.2  & 92.6 & 93.6 & 93.1 & 81.5 & 13.8 & 23.7 & 83.2 & 100  & 90.8 & 60.0   & 51.3 & 55.3 & 26.2 & 21.3  & 23.5 & 36.4 & 52.0   & 42.9 & 74.5 & 72.1 & 73.3 \\ \hline
\textsf{Ditto}            & 100 & 86.4  & 92.7  & 90.2 & 90.3 & 90.2 & 79.3 & 14.5 & 24.5 & 88.0 & 88.5 & 88.3 & 56.8 & 47.1 & 51.5 & 29.5 & 31.3  & 30.3 & 34.7 & 50.5 & 41.1 & 74.7 & 87.2 & 80.5 \\
\textsf{DADER}            & 81.8  & 81.8  & 81.8  & 81.5 & 91.4 & 86.2 & 98.4 & 37.7 & 54.6 & 87.6 & 96.2 & 91.7 & 15.0 & 87.4 & 25.6 & 11.4 & 100 & 20.5 & 26.1 & 64.6 & 37.2 & 54.2 & 92.7 & 68.4 \\
\textsf{Rotom}            & 100 & 77.3  & 87.2  & 89.2 & 94.3 & 91.7 & 83.3 & 15.7 & 26.5 & 95.8 & 88.0 & 91.7 & 68.0 & 54.6 & 60.5 & 43.6 & 34.1  & 38.3 & 51.9 & 45.5 & 48.5 & 76.7 & 78.5 & 77.6 \\ \hline
\textsf{TDmatch}          & 56.4  & 100 & 72.1  & 93.7 & 42.0 & 58.0 & 97.2 & 88.1 & \textbf{92.4} & 97.5 & 85.8 & 91.3 & 69.0 & 10.4 & 18.0 & 42.3 & 14.2  & 21.3 & 80.2 & 47.3 & 59.5 & 72.8 & 73.0 & 72.9 \\
\textsf{TDmatch*} & 10.0 & 4.6 & 6.3 & 80.2 & 87.3 & 83.6 & 37.5 & 18.9 & 25.1 & 66.5 & 77.1 & 71.4 & 42.7 & 30.2 & 35.4 & 32.0 & 23.2 & 26.9 & 48.6 & 40.3 & 44.1 & 51.0 & 51.0 & 51.0 \\ \hline
\textsf{PromptEM}         & 100 & 100 & \textbf{100} & 94.2 & 94.1 & \textbf{94.2} & 93.9 & 57.9 & 71.6 & 91.4 & 98.9 & \textbf{95.0} & 80.6 & 65.5 & 72.3 & 44.9 & 37.9  & \textbf{41.1} & 61.2 & 61.5 & \textbf{61.4} & 78.8 & 89.9 & 84.0 \\
\textsf{PromptEM w/o PT}  & 100 & 95.5  & 97.7  & 90.5 & 94.8 & 93.3 & 39.5 & 52.2 & 45.0 & 83.8 & 53.6 & 65.3 & 56.3 & 55.3 & 55.8 & 31.5 & 19.0  & 23.7 & 22.1 & 55.9 & 31.7 & 79.3 & 85.1 & 82.1 \\
\textsf{PromptEM w/o LST} & 100 & 100 & \textbf{100} & 94.2 & 94.1 & \textbf{94.2} & 91.7 & 27.7 & 42.5 & 91.4 & 98.9 & \textbf{95.0} & 73.4 & 59.1 & 65.5 & 35.8 & 31.8  & 33.7 & 58.0 & 60.4 & 59.2 & 76.4 & 90.4 & 82.8 \\
\textsf{PromptEM w/o DDP} & 100 & 100 & \textbf{100} & 93.0 & 94.8 & 93.9 & 80.5 & 57.2 & 66.9 & 92.1 & 95.1 & 93.6 & 79.0 & 68.9 & \textbf{73.6} & 47.3 & 33.2  & 39.0 & 55.5 & 63.3 & 59.2 & 85.0 & 88.1 & \textbf{86.5} \\ \hline
\end{tabular}}
\vspace{-3.5mm}
\end{table*}

\begin{table*}[h] \small
\centering
\caption{\rev{Results of all the methods under the extremely challenging low-resource setting.}}
\vspace{-4.5mm}
\label{table:challenging}
\setlength{\tabcolsep}{0.8mm}{
\begin{tabular}{|c|ccc|ccc|ccc|ccc|ccc|ccc|ccc|ccc|}
\hline
\makebox[0.1334\textwidth][c]{\multirow{2}{*}{\textbf{Methods}}} & \multicolumn{3}{c|}{\textbf{REL-HETER}} & \multicolumn{3}{c|}{\textbf{SEMI-HOMO}} & \multicolumn{3}{c|}{\textbf{SEMI-HETER}} & \multicolumn{3}{c|}{\textbf{SEMI-REL}} & \multicolumn{3}{c|}{\textbf{SEMI-TEXT-c}} & \multicolumn{3}{c|}{\textbf{SEMI-TEXT-w}} & \multicolumn{3}{c|}{\textbf{REL-TEXT}} & \multicolumn{3}{c|}{\textbf{GEO-HETER}} \\
                        & P        & R        & F        & P        & R        & F        & P         & R        & F        & P        & R        & F       & P         & R         & F        & P         & R         & F        & P        & R        & F       & P        & R        & F        \\ \hline
\textsf{DeepMatcher}             & 32.3     & 45.5     & 37.7     & 46.9     & 22.5     & 30.4     & 28.6      & 40.3     & 33.4     & 43.0       & 92.9     & 58.8    & 0.0         & 0.0         & 0.0        & 28.6      & 1.0         & 1.8      & 9.1      & 0.2      & 0.4     & 29.1     & 100      & 45.1     \\
\textsf{BERT}                    & 100      & 95.5     & 97.7     & 85.9     & 82.4     & 85.8     & 78.8      & 16.4     & 27.1     & 97.0       & 71.6     & 82.4    & 21.2      & 31.4      & 25.3     & 18.4      & 8.5       & 11.7     & 18.6     & 9.9      & 12.9    & 45.8     & 78.5     & 57.8     \\
\textsf{SentenceBERT}            & 95.5     & 95.5     & 95.5     & 86.8     & 73.3     & 79.5     & 100       & 14.5     & 25.3     & 71.0       & 74.9     & 72.9    & 20.8      & 13.8      & 16.6     & 23.1      & 8.5       & 12.5     & 15.5     & 4.1      & 6.4     & 53.8     & 64.8     & 58.8     \\ \hline
\textsf{Ditto}                   & 100      & 81.8     & 90.0       & 80.9     & 81.8     & 81.4     & 95.2      & 12.6     & 22.2     & 78.7     & 95.1     & 86.1    & 13.9      & 100       & 24.3     & 12.6      & 69.7      & 21.4     & 18.0       & 99.3     & 30.4    & 33.0       & 85.4     & 47.6     \\
\textsf{DADER}                   & 88.9     & 72.7     & 80.0       & 75.9     & 86.6     & 80.9     & 47.1      & 65.0       & 54.6     & 94.0       & 86.3     & 90.0      & 41.7      & 0.9       & 1.7      & 12.4      & 7.1       & 9.0        & 60.2     & 11.9     & 19.9    & 63.0       & 87.9     & 73.4     \\
\textsf{Rotom}                   & 100      & 95.5     & 97.7     & 80.1     & 93.5     & 86.2     & 77.8      & 17.6     & 28.7     & 96.3     & 85.8     & 90.8    & 23.5      & 31.4      & 26.9     & 22.0        & 5.2       & 8.4      & 19.8     & 23.9     & 21.6    & 72.0       & 78.3     & 75.0       \\ \hline
\textsf{TDmatch}                 & 56.4     & 100      & 72.1     & 93.7     & 42.0       & 58.0       & 97.2      & 88.1     & \textbf{92.4}     & 97.5     & 85.8     & 91.3    & 69.0        & 10.4      & 18.0       & 42.3      & 14.2      & 21.3     & 80.2     & 47.3     & \textbf{59.5}    & 72.8     & 73.0       & 72.9     \\
\textsf{TDmatch*}                & 11.1     & 9.1      & 10.0       & 37.8     & 27.8     & 32.0       & 37.8      & 17.6     & 24.0       & 47.8     & 75.4     & 58.5    & 16.3      & 5.2       & 7.9      & 19.0        & 7.1       & 10.3     & 20.0       & 14.4     & 16.8    & 36.7     & 33.6     & 35.1     \\ \hline
\textsf{PromptEM}                & 100      & 100      & \textbf{100}      & 86.1     & 92.2     & \textbf{89.0}      & 93.9      & 28.9     & 44.2     & 94.0       & 94.5     & \textbf{94.3}    & 40.8      & 29.0        & \textbf{33.9}     & 15.7      & 46.9      & \textbf{23.6}     & 26.5     & 50.2     & 34.7    & 78.0       & 81.9     & \textbf{79.9}     \\ \hline
\end{tabular}
}
\vspace{-4.5mm}
\end{table*}

\noindent \textbf{Implementation details.} 
We implement \textsf{PromptEM} in PyTorch~\cite{paszke2019pytorch}, the Transformers library \cite{wolf2019huggingface} and the OpenPrompt library \cite{ding2021openprompt}. We use RoBERTa-base \cite{liu2019roberta} as the backbone structure of our model in all the experiments. Unless particularly specified, the experimental results are conducted under the low-resource setting shown in Table \ref{table:dataset}. We further apply the half-precision floating-point (fp16) optimization to save the GPU memory usage and running time. In all the experiments, the max sequence length is set to 512; the learning rate is set to 2e-5; the batch size is set to 32; the number of iterations for self-training is set to 1; and the number of passes for MC-Dropout is set to 10. We use AdamW as the optimizer for training, fix the epochs of training the teacher model to 20, and set the epochs of training the student model to 30. We prune the train set for every 8 epochs. We tune the hyper-parameters by doing a grid search and selecting the one with the best performance. Specifically, the continuous template is selected from \{$\operatorname{T_1(\cdot)}$, $\operatorname{T_2(\cdot)}$\}, $u_r$ is selected from \{0.05, 0.10, 0.15, 0.20, 0.25\}, and $e_r$ is selected from \{0.1, 0.2, 0.3, 0.4, 0.5\}. We select the epoch with the highest F1-score on the validation set, and report the values of precision, recall, and F1-score on the test set. All the experiments are conducted on a machine with an Intel Xeon Silver 4216 CPU, an NVIDIA A100 GPU and 512GB memory.
\rev{We use the same serializing method as \textsf{PromptEM} to implement each baseline method and report the performance under their optimal settings. 
We present the implementation of baselines in our online version.}

\noindent \textbf{Evaluation metrics.}
Following related studies \cite{li2020deep, ge2021collaborem, tu2022domain}, we employ three widely-used classification metrics, namely, precision (P), recall (R), F1-score (F).
\vspace{-2.8mm}
\subsection{Main Results (RQ1)}
\noindent \textbf{Results under the default low-resource setting.}
We first verify the performance under low-resource setting of \textsf{PromptEM} using the above \rev{eight} baselines. The benchmark results of all methods across the datasets are reported in Table \ref{table:low}.
\textsf{DeepMatcher} achieves the worst performance, since it does not leverage the recent advances in pre-trained LMs. Existing low-resource EM approaches (i.e., \textsf{Ditto}, \textsf{DADER}, and \textsf{Rotom}) achieve relatively poor performance, because the GEM problem is more intractable than EM (e.g., heterogeneous tables). 
In particular, \textsf{TDmatch} is not stable across different datasets due to the absence of label guidance, which can achieve the best F1-score on SEMI-HETER but only 18.0 F1-score on SEMI-TEXT-c. 
\rev{\textsf{TDmatch} outperform other LM-based approaches on SEMI-HETER. The reason is that SEMI-HETER has lots of numeric attributes, i.e., 53\% attribute values are digits. It is well known that LMs are not good at understanding digits \cite{Wallace2019DoNM}.}
Also, we find that the scalability of \textsf{TDmatch} is extremely poor, which will be confirmed in Section~\ref{exp:efficiency}.  
\rev{We can observe that \textsf{TDmatch} performs better than \textsf{TDmatch*} in most cases.
This is because \textsf{TDmatch} is specifically designed for unsupervised learning, not necessarily suitable for supervised learning. Besides, we have a similar finding as those studies \cite{li2019fine, krueger2017deep}: BERT can be generalized to the specific domain, and hence, BERT based methods (including \textsf{PromptEM}) achieve better performance on GEO-HETER.}

\begin{figure*}[ht]
\centering
\includegraphics[width=0.9\textwidth]{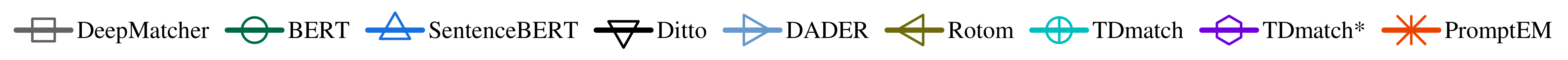}\\
\vspace*{-4mm}
\subfigure[REL-HETER]{
 \includegraphics[width=1.6in]{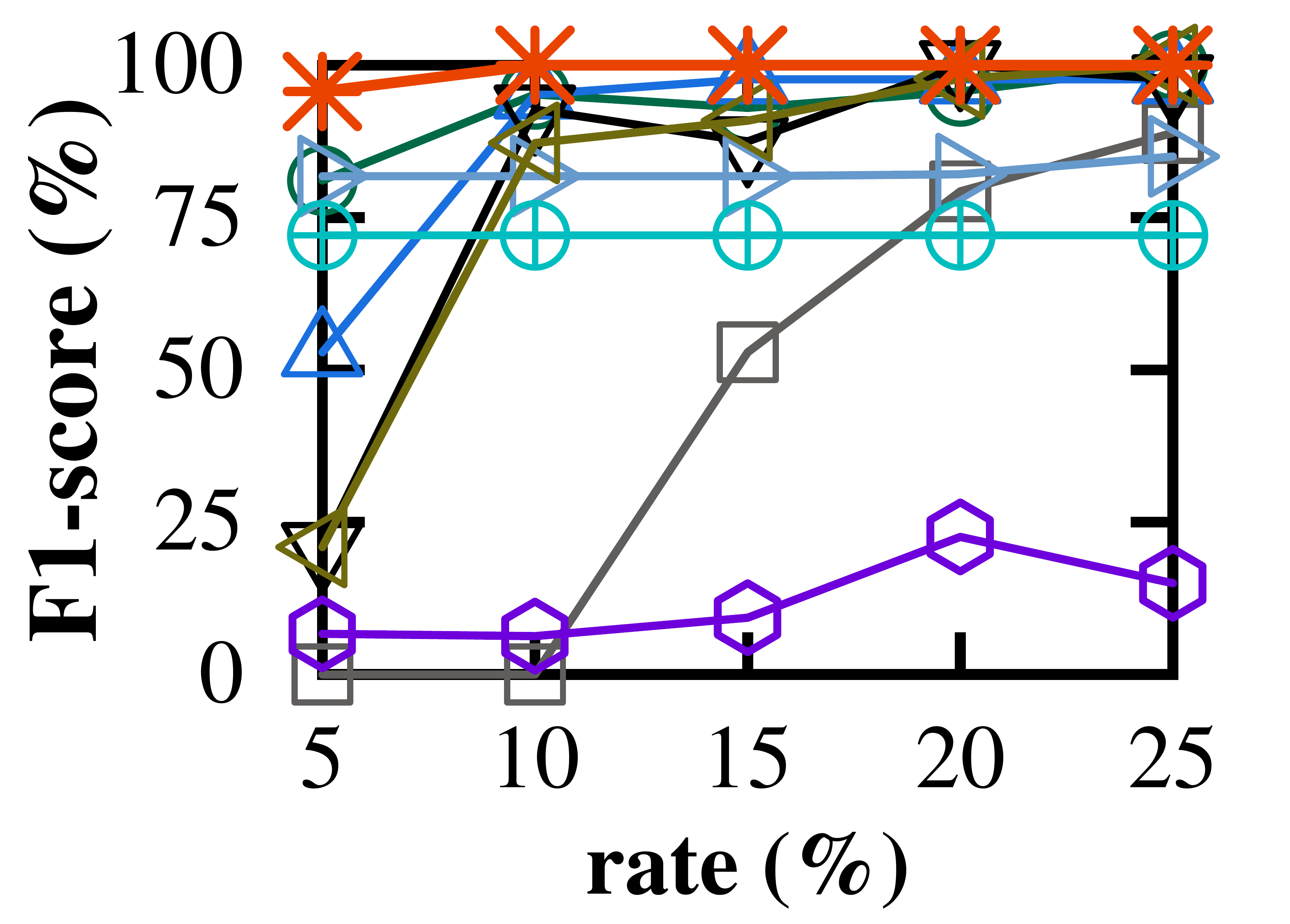}
}
\subfigure[SEMI-HOMO]{
 \includegraphics[width=1.6in]{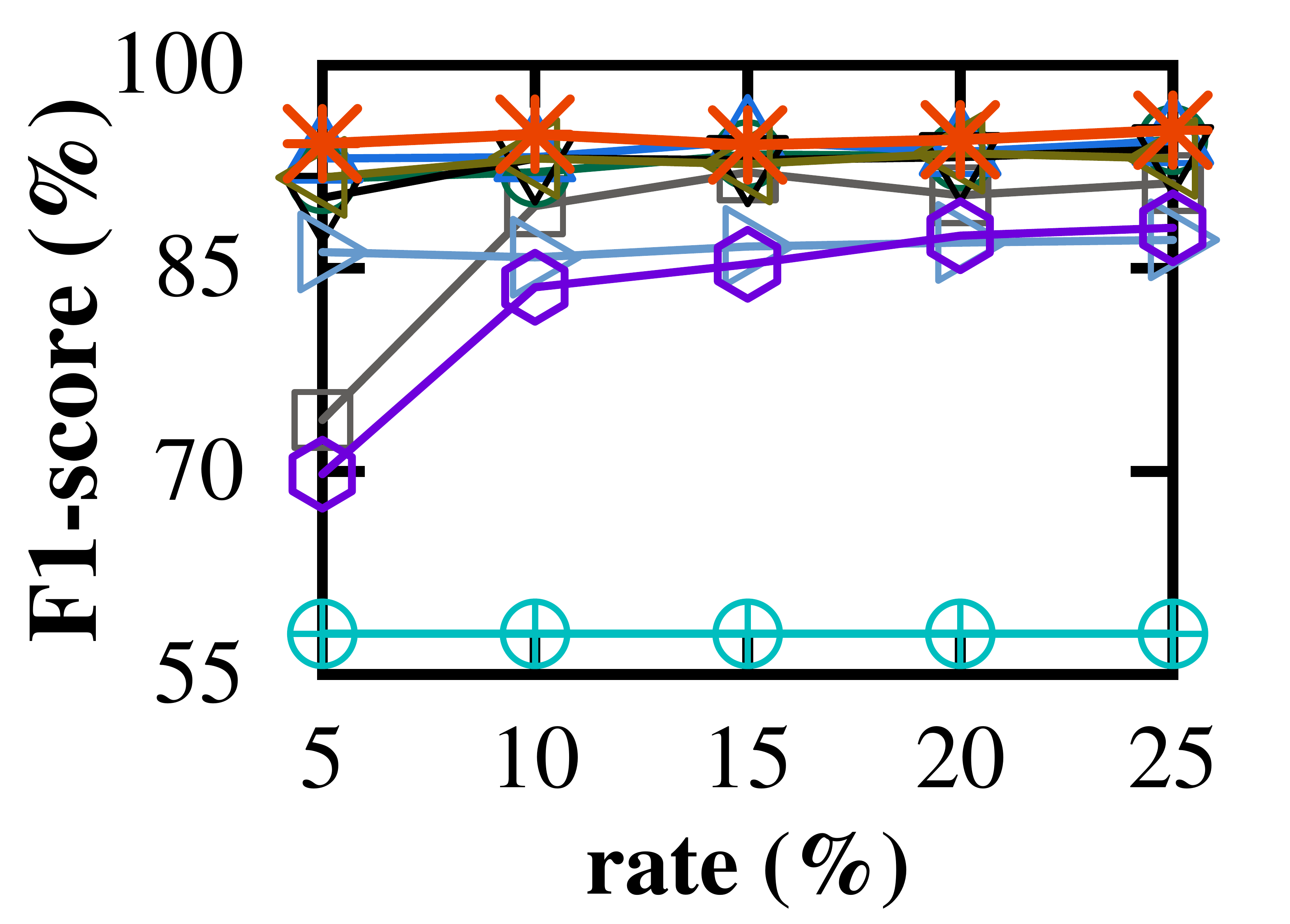}
}
\subfigure[SEMI-HETER]{
 \includegraphics[width=1.6in]{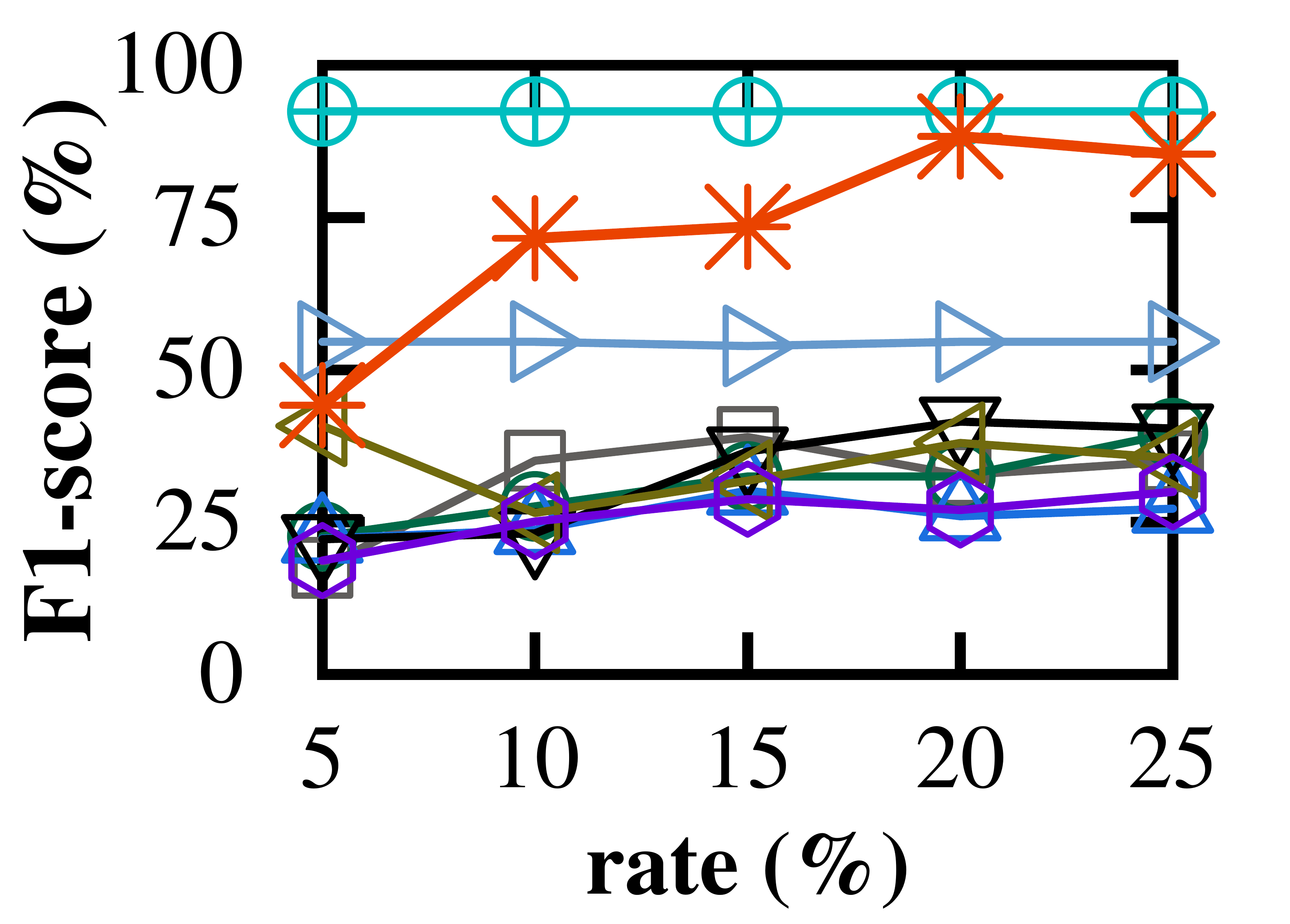}
}
\subfigure[SEMI-REL]{
 \includegraphics[width=1.6in]{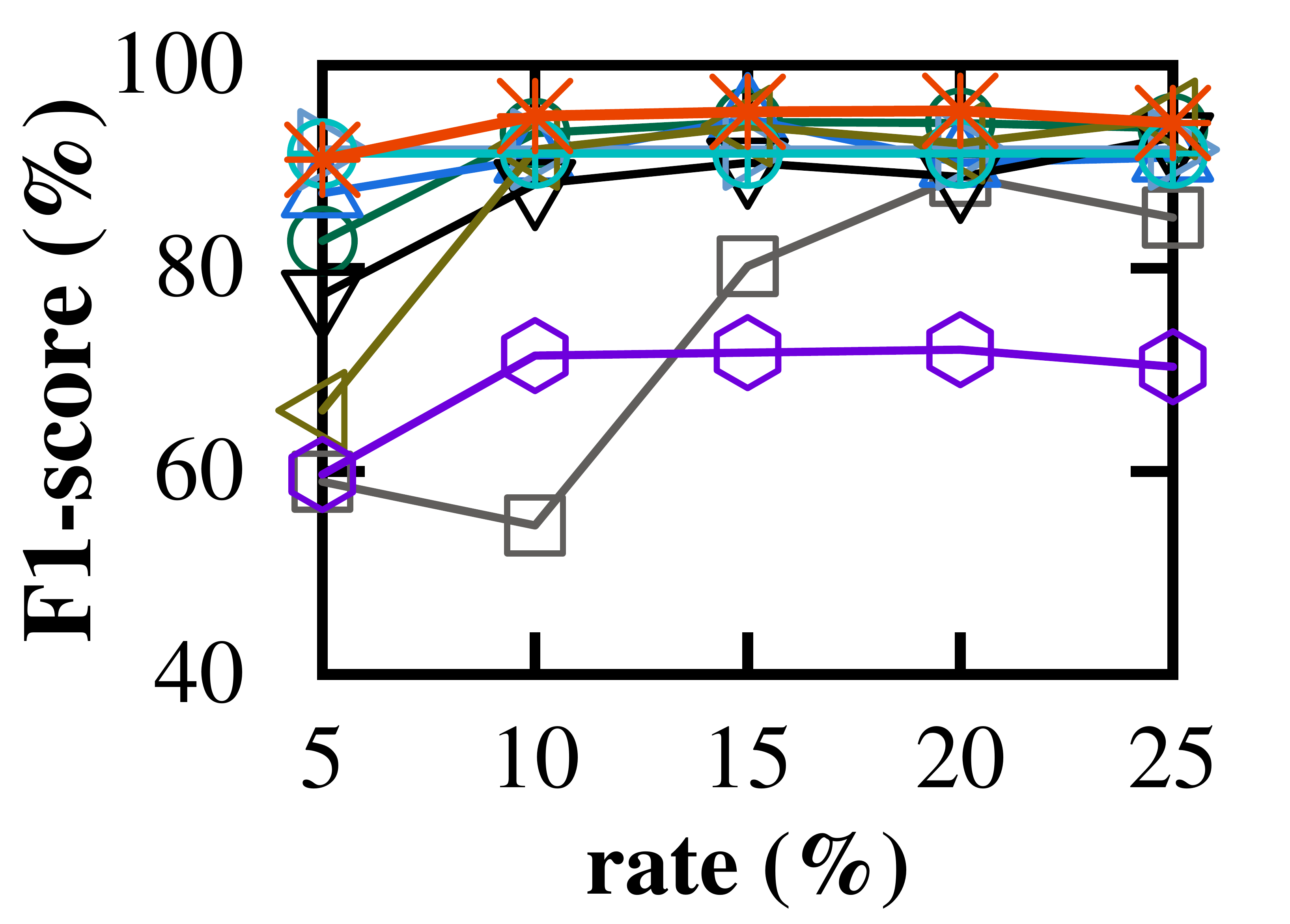}
}\\ \vspace{-4mm}
\subfigure[SEMI-TEXT-c]{
 \includegraphics[width=1.6in]{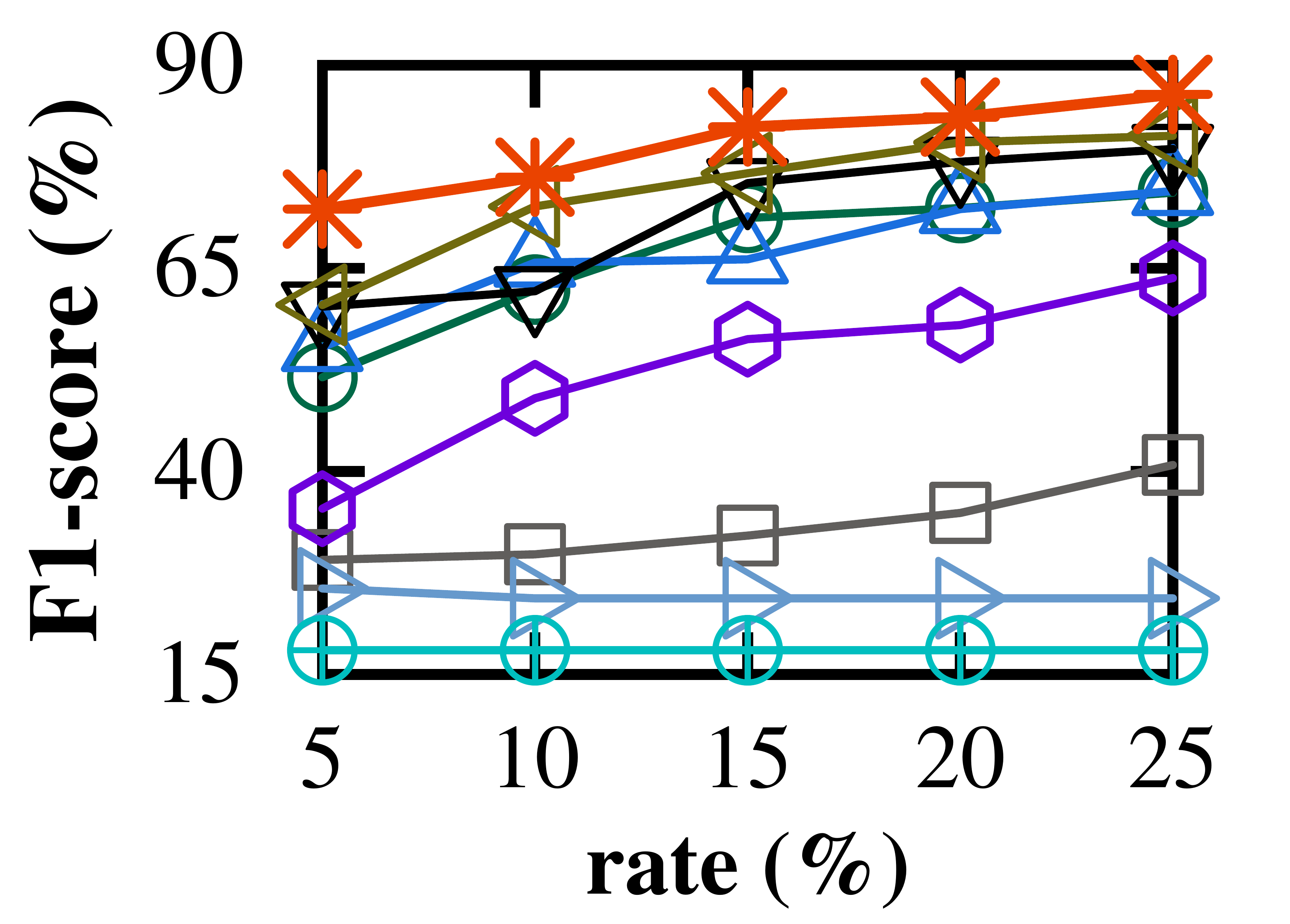}
}
\subfigure[SEMI-TEXT-w]{
 \includegraphics[width=1.6in]{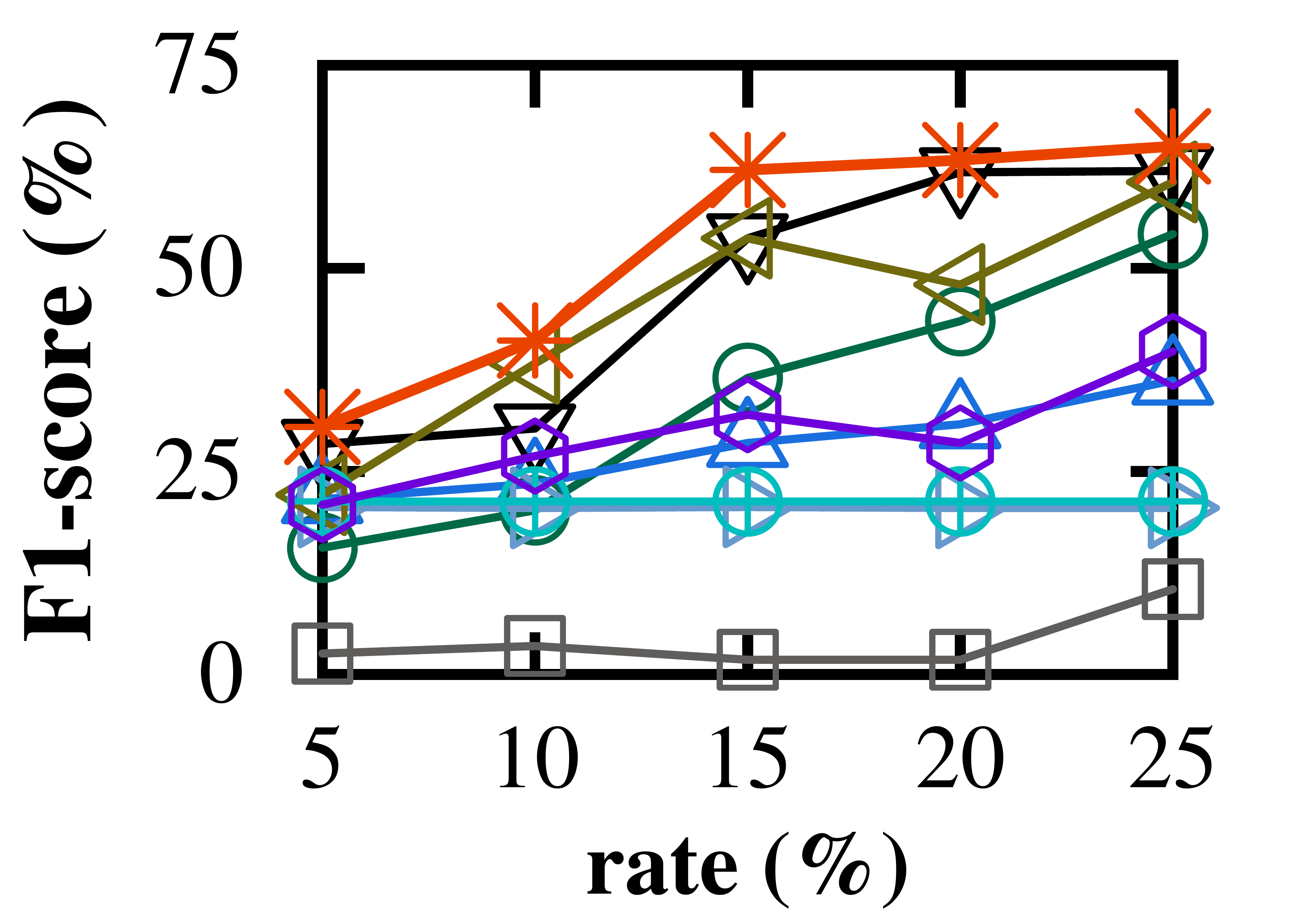}
}
\subfigure[REL-TEXT]{
 \includegraphics[width=1.6in]{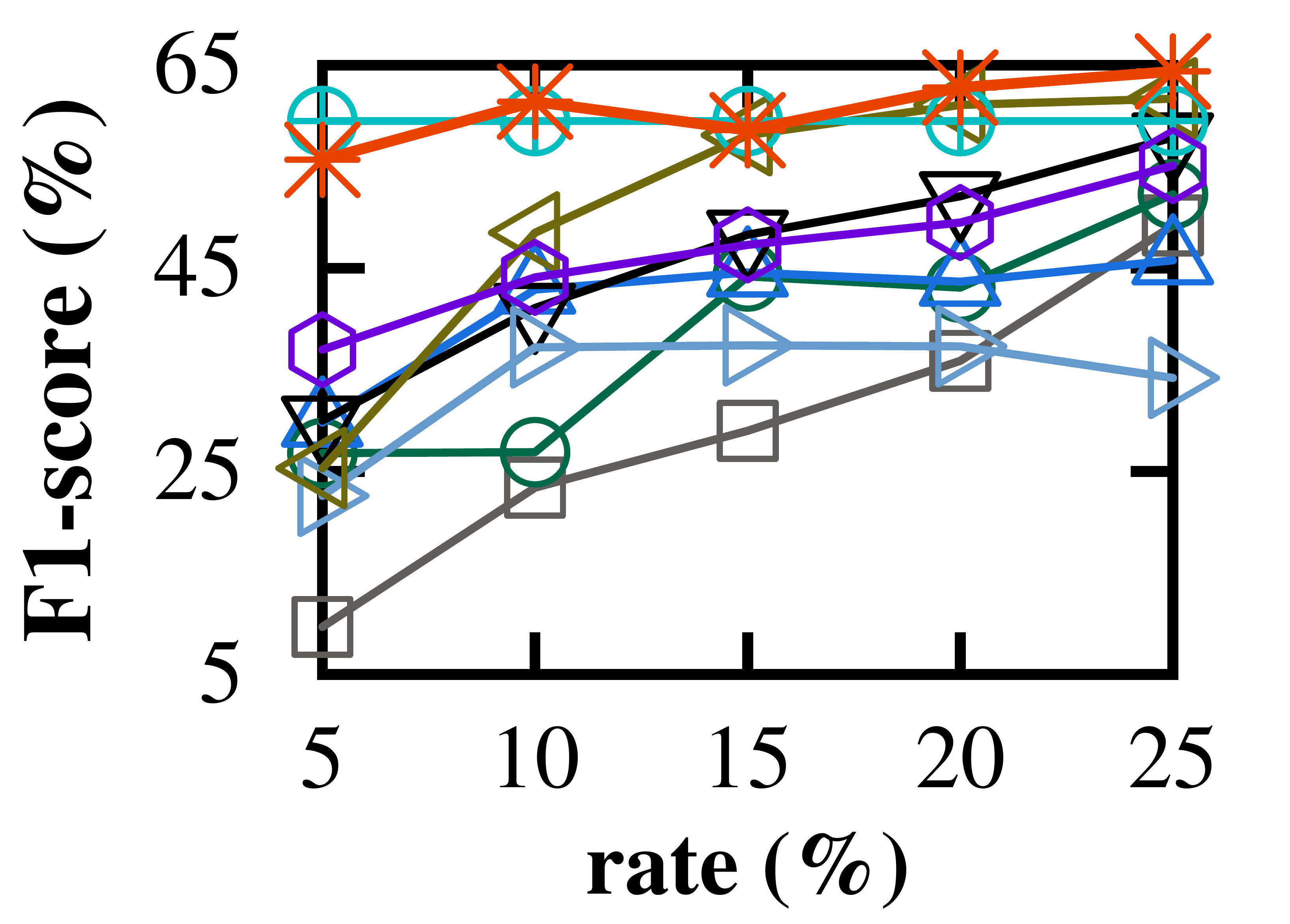}
}
\subfigure[GEO-HETER]{
 \includegraphics[width=1.6in]{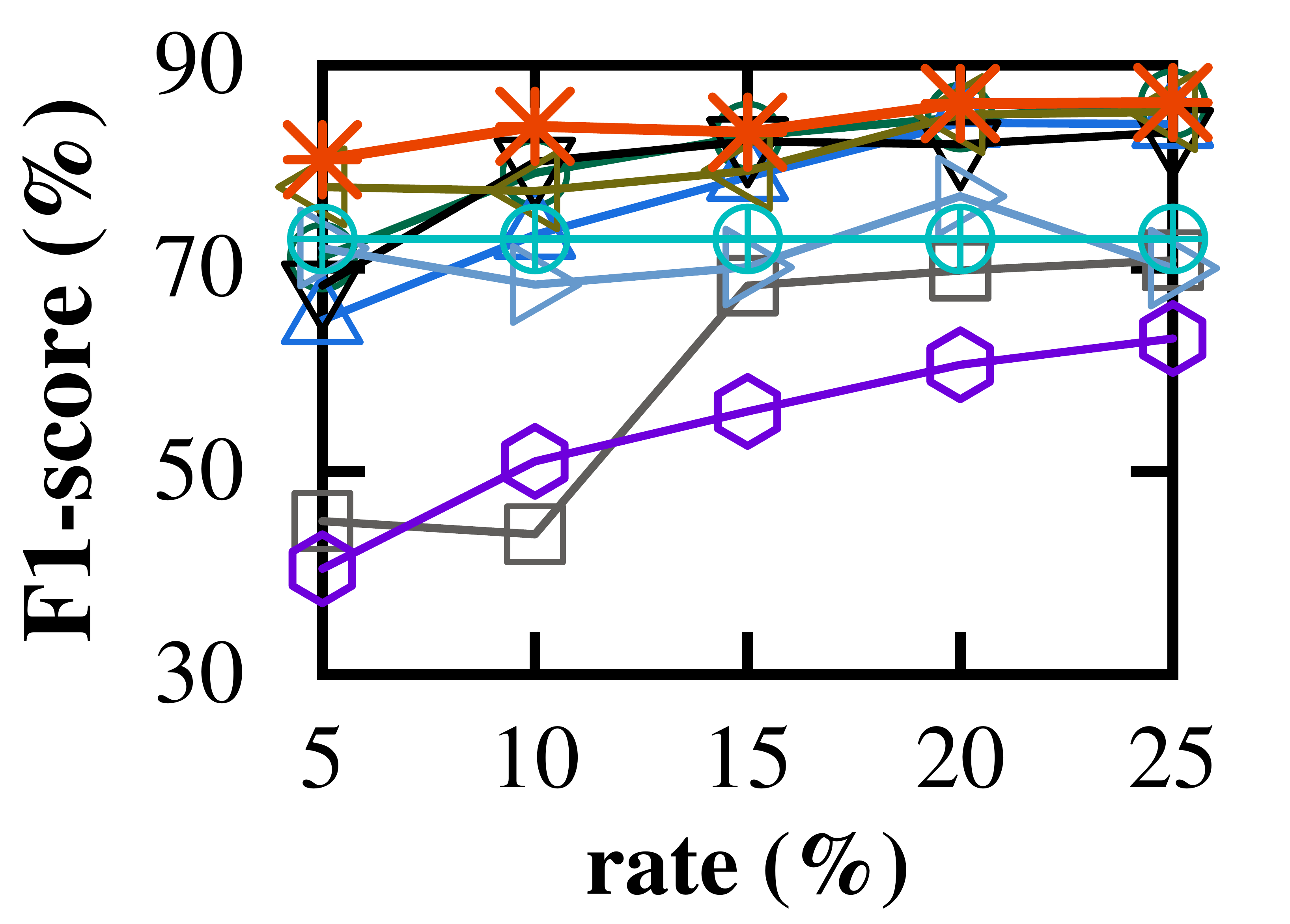}
}\\
\vspace*{-5.5mm}
\caption{\rev{Comparison results under different low-resource settings (\%).}}
\label{fig:different_low}
\vspace*{-6mm}
\end{figure*}

\vspace{1.75mm}
\noindent \textbf{Effectiveness to different low-resource settings.}
We reduce the training rate from 25\% to 5\% to see the performance under different low-resource settings. Experimental results are depicted in Figure \ref{fig:different_low}. We observe that \textsf{PromptEM} achieves SOTA performance in most cases, while \textsf{TDmatch} and \textsf{DADER} achieve unstable results across different datasets due to lacking the guidance of labels and the heterogeneity of datasets.
\rev{We also evaluate methods in a more challenging setting, i.e., the number of available data for training is only 80 for all the datasets.
This setting is extremely challenging for supervised methods, e.g., only using 0.46\% labeled examples on SEMI-HOMO.
As shown in Table \ref{table:challenging}, \textsf{PromptEM} achieves SOTA performance on most datasets, which demonstrates the great robustness of \textsf{PromptEM} compared to baselines.
Moreover, it also shows the outstanding scalability of \textsf{PromptEM}, which achieves considerable performance only using a small number of labeled data examples.
}

\rev{Overall, our \textsf{PromptEM} is superior to all the baselines in almost all the cases under various low-resource settings.
As mentioned in Challenge \uppercase\expandafter{\romannumeral 1}, there is a significant gap between objective forms in pre-training and fine-tuning. 
This gap hinders the transfer and adaptation of knowledge in LMs for GEM tasks, which restricts taking full advantage of knowledge in LMs.
Prompt-tuning is a new promising paradigm in natural language processing, and is able to bridge the gap of objective forms between pre-training and fine-tuning~\cite{liu2021pre, han2021pre}.
Thus, we can stimulate the rich knowledge distributed in LMs through designing GEM-specific prompt-tuning \cite{liu2021pre, han2021pre}.
Recently, prompt-tuning has been applied to lots of tasks such as machine translation, information extraction, question answering, and so on \cite{liu2021pre}.
In summary, prompt-tuning has the potential to outperform fine-tuning for those tasks based on LMs.
} 
% Our results suggest that the prompt-tuning paradigm has the potential to outperform the fine-tuning paradigm in data management tasks, much as it has revolutionized NLP. 
\vspace{-4mm}
\subsection{Ablation Study (RQ2)}
\label{exp:abla}
Next, we study the effectiveness of each module in \textsf{PromptEM} (i.e., prompt-tuning (PT), lightweight self-training (LST), dynamic data pruning (DDP)) by comparing \textsf{PromptEM} with its variants without the key module. The results are listed in Tables \ref{table:low}.

\noindent \textbf{PromptEM vs. PromptEM w/o PT.} PromptEM w/o PT denotes that we fine-tune the LM instead of prompt-tuning. It is observed that the use of prompt-tuning contributes to a large portion of the performance gain. The F1-score drops \rev{15.7\%} on average under the low-resource setting. This confirms that prompt-tuning greatly helps to stimulate the rich knowledge distributed in the LM.

\noindent \textbf{PromptEM vs. PromptEM w/o LST.} We use LST to boost the performance under low-resource settings. We can observe that LST can bring performance improvement in most cases. For example, LST brings 6.8\% improvement on SEMI-TEXT-c. Also notice that LST brings relatively low improvement on some datasets. This is attributes to the nature of the datasets, as it is relatively much easier for \textsf{PromptEM} to achieve the extremely high performance, e.g., 100\% F1-score on REL-HETER. 
% \rev{Since LST is general enough to incorporate with other approaches, it is possible to be widely used in practical low-resource GEM applications.}

\noindent \textbf{PromptEM vs. PromptEM w/o DDP.} We can observe that DDP can prune useless training data without sacrificing test accuracy. It is worth noting that DDP can prune training data while slightly improving test accuracy in some datasets. This is because DDP makes the model focus on important and useful training data.

\vspace{-3mm}
\subsection{Efficiency Analysis (RQ3)}
\label{exp:efficiency}
\rev{We further explore the efficiency of our proposed \textsf{PromptEM} in terms of training time and memory usage, and the results are presented in Table \ref{table:time}.} Since it is common for methods to use a similar strategy for evaluating the GEM results in the test set, we do not report the test time of every evaluated approach.
\textsf{SBERT} denotes \textsf{SentenceBERT}, and \textsf{PromptEM-} represents \textsf{PromptEM} without dynamic data pruning.

\noindent \textbf{PromptEM vs. best baselines.}
Due to the limitation of space, we report \textsf{PromptEM} with the other evaluated approaches that achieve the best quality of GEM results in the corresponding categories, i.e., the normal EM method \textsf{SBERT}, the low-resource EM approach \textsf{Rotom}, and the unsupervised matching method \textsf{TDmatch}. 
\rev{We report the GPU memory for the methods running on GPU and the CPU memory for the method (i.e., \textsf{TDmatch}) running on CPU, respectively.}
As observed, \textsf{PromptEM} needs more training time than \textsf{SBERT} to obtain the SOTA results. This demonstrates a trade-off between the effectiveness and efficiency of the GEM problem. To sum up, it is significant that spending a relatively longer time in achieving better matching results. \textsf{Rotom} requires two-stage training, incurring a long training process. 
\rev{\textsf{SBERT}, \textsf{Rotom}, and \textsf{PromptEM} need similar memory usage since they are all based on LMs. We would like to emphasize that \textsf{TDmatch} needs too much training time and memory usage, especially on relatively large datasets (e.g., 120.3 hours and 131.5 Gigabytes on SEMI-REL), which is very costly in real-world applications.}

\noindent \textbf{PromptEM vs. PromptEM-.}
We also compare \textsf{PromptEM} with \textsf{PromptEM-} to evaluate the efficiency of dynamic data pruning.
It is observed that DDP greatly helps to reduce the training time, i.e., reduce \rev{26.1\%} time on average. This is because the proposed MC-EL2N is able to quantify useless training data effectively. 
\rev{Meanwhile, DDP does not bring extra memory usage as it does not require any new model parameters.}
As analyzed in Section \ref{exp:abla}, DDP does not hurt the performance. This further demonstrates that \textsf{PromptEM} is effective and efficient in solving the GEM problem.

\begin{table}[] \small
\centering
\caption{\rev{Efficiency comparison between \textsf{PromptEM} and its competitors, including the running time and memory usage. "s" denotes seconds, "m" denotes minutes, "h" denotes hours, and "G" represents gigabytes. Due to the limitation of space, we use the abbreviations of datasets.}}
\label{table:time}
\vspace{-4mm}
\setlength{\tabcolsep}{0.4mm}{
\begin{tabular}{|c|cc|cc|cc|cc|cc|}
\hline
\multirow{2}{*}{Datasets} & \multicolumn{2}{c|}{\textsf{SBERT}} & \multicolumn{2}{c|}{\textsf{Rotom}} & \multicolumn{2}{c|}{\textsf{TDmatch}} & \multicolumn{2}{c|}{\textsf{PromptEM-}} & \multicolumn{2}{c|}{\textsf{PromptEM}} \\
                          & T.         & M.        & T.         & M.        & T.          & M.         & T.           & M.          & T.          & M.         \\ \hline
R-H                 & 28.8s        & 22.9G       & 35.4s        & 32.8G       & 14.0m         & 6.2G        & 38.3s          & 27.4G         & 26.6s         & 27.4G        \\
S-HO                 & 2.4m         & 29.5G    & 8.2m         &    32.8G         & 51.0h         & 41.4G             & 11.5m          & 29.2G              & 7.4m          & 29.2G             \\
S-HE                & 43.4s        &  24.9G           & 1.8m         & 35.8G            & 102.8h        &   105.3G           & 1.6m          & 29.0G              & 1.5m         &   29.0G           \\
S-R                  & 50.1s        & 35.8G            & 2.6m         &    32.8G         & 120.3h        &     131.5G         & 1.4m          &  30.6G             & 1.1m         & 30.6G             \\
S-T-c               & 2.0m         &  36.3G           & 20.2m        &    32.8G         & 10.7h         & 25.4G             & 20.8m          &    29.2G           & 11.5m         &   29.2G           \\
S-T-w               & 1.8m         &  35.6G           & 11.2m        &    32.8G         & 2.2h          & 12.9G             & 6.1m           &  29.2G             & 5.3m          &  29.2G            \\
R-T                  & 2.3m         &  34.4G           & 11.1m        &    29.7G         & 5.6h          &     50.5G         & 8.1m           &    30.6G           & 5.6m          &   30.6G           \\
G-H                 & 49.2s             & 27.9G            & 1.3m             &  32.8G           &    36.2m           & 16.7G             & 6.2m               & 30.6G              & 4.6m              & 30.6G             \\ \hline
\end{tabular}}
\vspace{-3.5mm}
\end{table}

\vspace{-3.2mm}
\subsection{Model Variants (RQ4)}
\label{sec:variants}
Finally, we investigate the performance of \textsf{PromptEM} using alternative modules by conducting the following experiments.

\noindent \textbf{Effect of template choices.} Designing prompt templates is a primary component of prompt-tuning. We verify the effect of different templates, i.e., continuous $\operatorname{T_1{(\cdot)}}$, hard-encoding $\operatorname{T_1{(\cdot)}}$, continuous $\operatorname{T_2{(\cdot)}}$ and hard-encoding $\operatorname{T_2{(\cdot)}}$. \rev{Their average F1-scores on all datasets are 74.4, 67.8, 77.0, and 74.5, respectively. Continuous templates achieve better performance than hard-encoding templates.}
This further validates the effectiveness of the proposed continuous templates, which can find better continuous prompts beyond the original vocabulary $\mathcal{V}$ of $\mathcal{M}$ could express.

\begin{table}[t] \small
\centering
\caption{\rev{Results of pseudo-label selection strategies.}}
\vspace{-4mm}
\label{table:pseudo}
\setlength{\tabcolsep}{2.41mm}{
\begin{tabular}{|c|cc|cc|cc|}
\hline
\multirow{2}{*}{Datasets} & \multicolumn{2}{c|}{Uncertainty} & \multicolumn{2}{c|}{Confidence} & \multicolumn{2}{c|}{Clustering} \\ \cline{2-7} 
            & TPR    & TNR    & TPR    & TNR    & TPR    & TNR    \\ \hline
REL-HETER   & 1      & 1      & 0.250   & 0.864 & 0.250   & 0.881 \\
SEMI-HOMO   & 1      & 0.998 & 0.197 & 0.803 & 0.193  & 0.815 \\
SEMI-HETER  & 1      & 0.963 & 0.979 & 0.985 & 0.350 & 0      \\
SEMI-REL    & 1      & 1      & 0.426 & 0.583 & 0.432 & 0.602 \\
SEMI-TEXT-c & 0.969 & 1      & 0.113 & 0.897 & 0.128 & 0.879 \\
SEMI-TEXT-w & 0.333 & 0.967  & 0.056 & 0.928 & 0.114 & 0.912 \\
REL-TEXT    & 0.910   & 0.966 & 0.194 & 0.820 & 0.148 & 0.846 \\
GEO-HETER    & 0.867   & 1  & 0.644  & 0.758 & 0.236  & 0.738  \\\hline
\end{tabular}}
\vspace{-3mm}
\end{table}

\noindent \rev{\textbf{Effect of label words choices.}
We compare our designed label words with a simple one (i.e., matched and mismatched).
Using continuous $\text{T}_1$ and $\text{T}_2$, our designed label words achieve +5.2\% and +9.4\% average F1-score improvements over the simple one, respectively.
This confirms the effectiveness of our designed label words.
In other words, considering the more general relationship between entities is beneficial to the predictions.}

\noindent \textbf{\rev{Pseudo-label} selection strategies.} We consider several pseudo-label selection strategies, including uncertainty \cite{rizve2021defense}, confidence, and clustering \cite{dopierre2020few}. We fix $u_r$ to 0.1 on all datasets. Similarly, confidence and clustering both select the samples whose scores are in the top 10\%. Following \cite{ge2021collaborem}, we use true-positive rate (TPR) and true-negative rate (TNR) to evaluate the quality of the pseudo-labels generated by different strategies. Formally, TPR represents the proportion of matched entity pairs that are correctly labeled, denoted as $\frac{TP}{TP+FN}$; TNR represents the proportion of mismatched pairs that are correctly labeled, denoted as $\frac{TN}{TN+FP}$. The results are reported in Table \ref{table:pseudo}. As expected, uncertainty can achieve state-of-the-art performance when generating pseudo-labels, e.g., TPR and TNR are \rev{0.88} and \rev{0.99} on average, respectively. It confirms the effectiveness of the uncertainty-aware pseudo-label selection strategy.

\vspace{-1mm}
\section{Related Work}
\label{sec:related_work}

\subsection{Entity Matching}
Entity Matching (EM) is one of the fundamental and significant tasks in data management. Many efforts have devoted to develop effective approaches for EM, including rule-based methods \cite{elmagarmid2014nadeef,singh2017synthesizing,wang2011entity}, crowdsourcing-based methods \cite{gokhale2014corleone, marcus2011human, wang2012crowder}, and traditional ML-based methods \cite{bilenko2003adaptive, cohen2002learning, konda2016magellan, sarawagi2002interactive}. Recently, deep learning has been used widely in EM, and achieved promising results. \textsf{DeepER} \cite{ebraheem2018distributed} uses deep neural networks to extract features of entity pairs, and then models EM as a binary classification task. \textsf{DeepMatcher} \cite{mudgal2018deep} systematically describes a DL architecture, and designs a space of DL solutions for EM. However, a lot of labeled training data are still needed for those DL-based approaches, which is extremely expensive in practice. To decrease the demand for high-quality training data,
\textsf{Ditto} \cite{li2020deep} applies pre-trained language models to EM, performing well with the help of some data augmentation (DA) techniques. \textsf{Rotom} \cite{miao2021rotom} effectively improves the performance of EM tasks via combining multiple DA operators. \textsf{DADER} \cite{tu2022domain} develops a framework that significantly advances EM by applying domain adaptation. Some other attempts have also been made to enhance the performance via information fusion \cite{yao2021interpretable}, active learning \cite{kasai2019low,nafa2022active}, and transfer learning \cite{zhao2019auto,loster2021knowledge,thirumuruganathan2018reuse}. Nonetheless, these methods only focus on EM tasks in low-resource scenarios but perform poorly on GEM tasks. \textsf{TDmatch} \cite{ahmadi2021unsupervised} first attempts to match textual content and structured data under an unsupervised setting.
\rev{However, it has one serious shortcoming: it is not scalable on large-scale datasets, which makes it hard to be used in practical scenarios.}
\vspace{1mm}
\subsection{Prompt-tuning}
Despite the success of fine-tuning pre-trained LMs \cite{devlin2018bert, liu2019roberta}, the huge objective form gap between pre-training and fine-tuning still hinders the full use of pre-trained knowledge for downstream tasks \cite{liu2021gpt, liu2021pre}. The birth of \textsf{GPT-3} \cite{brown2020language} is the seminal work that stimulates the development of prompt-tuning, which applies hand-encoding prompts for tuning and achieves impressive performance on various tasks, especially under the low-resource settings. Following \textsf{GPT-3}, many hand-encoding prompts \cite{liu2021gpt, ding2021prompt} are widely explored. Recently, automatic prompt search \cite{shin2020autoprompt} and continuous prompts \cite{liu2021gpt, han2021ptr} are proposed to avoid labor-intensive prompt design and enhance the expressiveness of the prompt. The burst of prompt-tuning has led to significant advancement in many areas such as natural language inference \cite{liu2021gpt} and entity typing \cite{ding2021prompt}. However, for the first time, we introduce prompt-tuning in EM for the better usage of pre-trained LMs. 
\rev{\textsf{PromptEM} provides a good connection to recent NLP advancements with applications to the data management task.}
\vspace{1.3mm}
\section{Conclusions}
\label{sec:conclusion}
In this paper, we study the problem of low-resource generalized entity matching via our presented \textsf{PromptEM}. 
\rev{For the first time, \textsf{PromptEM} introduces prompt-tuning to cast GEM as a cloze-style task to bridge the gap between pre-training and fine-tuning. 
It is non-trivial as prompt templates are not generic and need to be specially designed for the task.
To this end, we design GEM-specific templates and label words set for the GEM problem.}
\rev{To further improve the performance in low-resource settings, we develop a generic lightweight self-training method using uncertainty, which is a nice combination and can be efficient and lightweight by dynamic data pruning.}
Extensive experimental results on \rev{eight} real-world datasets with different structures demonstrate the superiority of \textsf{PromptEM} compared with the state-of-the-art approaches. In the future, we plan to explore a general prompt-tuning method to support more data management tasks (e.g., data cleaning), advancing the usage of LMs in the database community.

\newpage
\balance
\clearpage

\bibliographystyle{ACM-Reference-Format}
\bibliography{ref}
\newpage
\appendix
\section*{Appendix}

\section{Results under Sufficient Resource Setting}
As shown in Table \ref{table:sufficient}, we observe that \textsf{PromptEM} achieves the best F1-score for all datasets. \textsf{PromptEM} achieves an average 88.9\% F1-score, which is +8.0\% relatively over the best baseline \textsf{Rotom}. This attributes to the power of prompt-tuning. As mentioned in Section \ref{sec:intro}, prompt-based tuning is able to bridge the gap of objective forms between pre-training and fine-tuning, which is good at stimulating the rich knowledge distributed in LMs.
PromptEM w/o PT denotes that we fine-tune the LM instead of prompt-tuning. It is observed that the use of prompt-tuning contributes to a large portion of the performance gain. The F1-score drops 5.2\% on average under the sufficient setting when fine-tuning.

\begin{figure}[h]
\centering
\includegraphics[width=.35\textwidth]{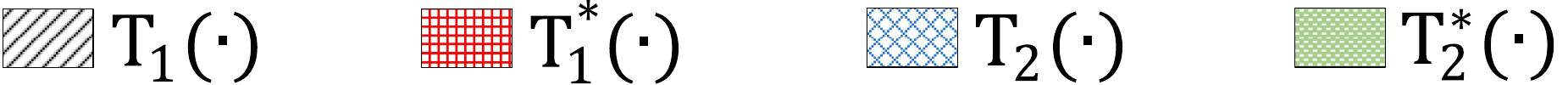}\\
\includegraphics[width=3.2in]{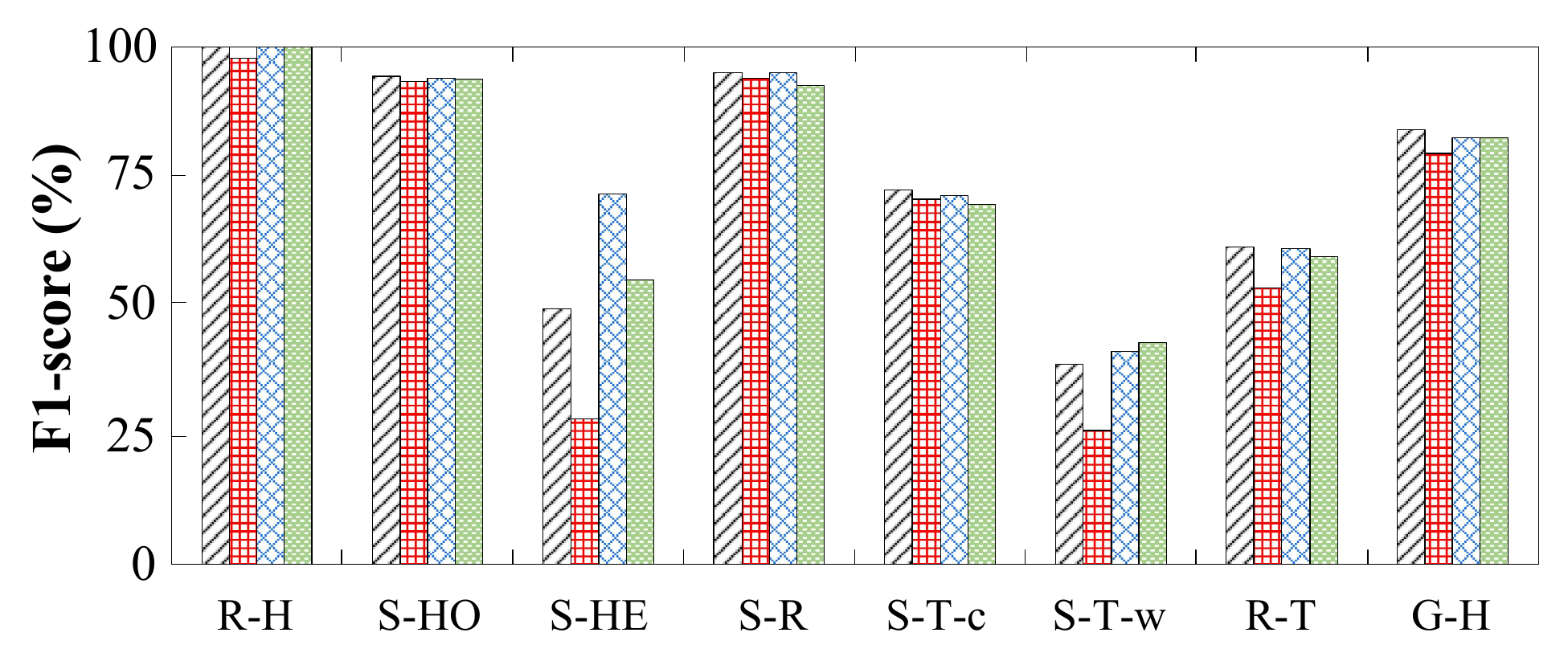}
\vspace{-4mm}
\caption{Effect of template choices. $\text{T}_1(\cdot)$ and $\text{T}_2(\cdot)$ denote the continuous templates. $\text{T}^{*}_1(\cdot)$ and $\text{T}^{*}_2(\cdot)$ represent the hard-encoding templates. We use the abbreviations of datasets.}
\label{fig:templates}
\vspace{-4mm}
\end{figure}

\begin{figure}[h]
\centering
\includegraphics[width=.4\textwidth]{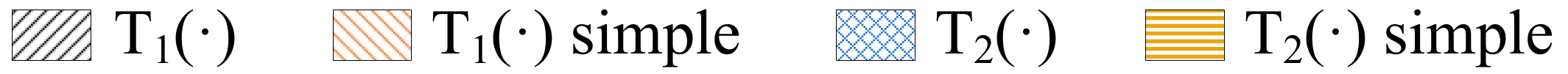}\\
\includegraphics[width=3.3in]{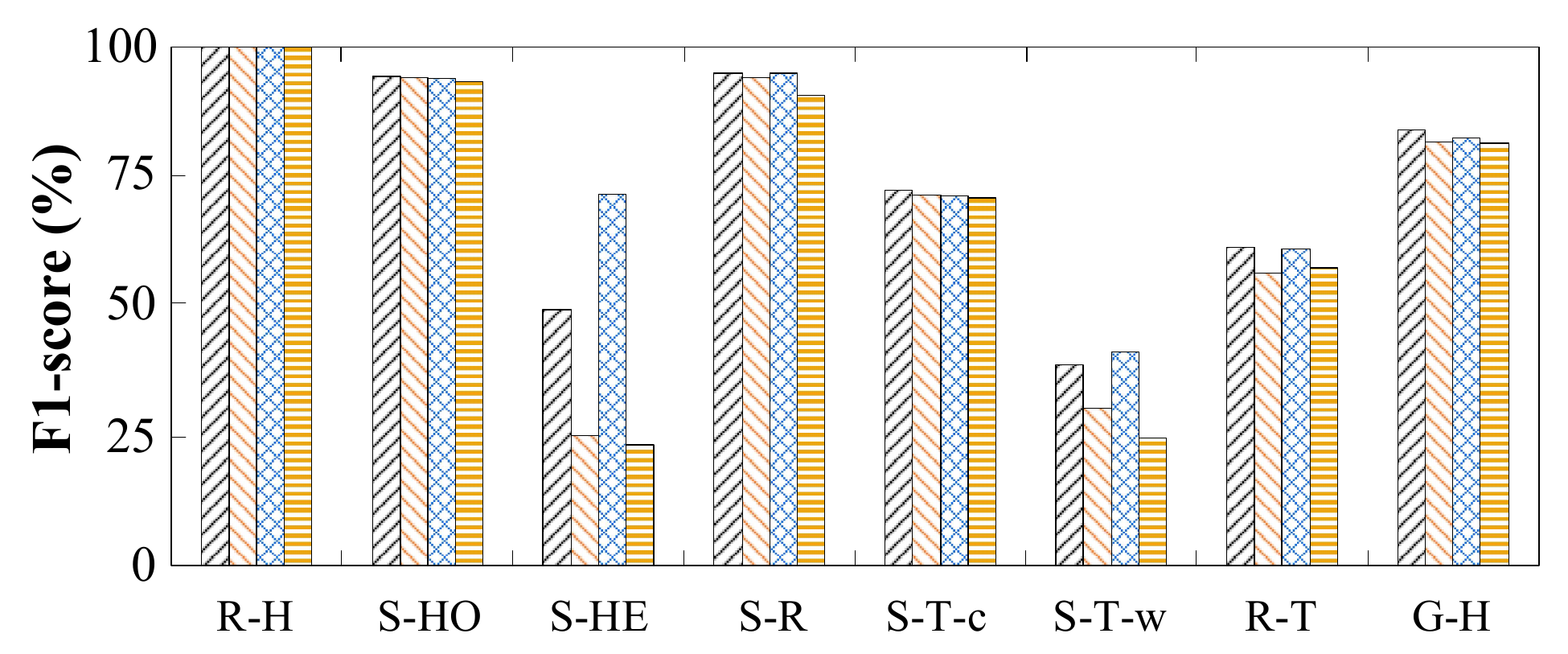}
\vspace{-4mm}
\caption{Effect of label words choices.
$\text{T}_1(\cdot)$ simple and $\text{T}_2(\cdot)$ simple denote using the simple label words, i.e., matched for "yes" and mismatched for "no".
We use the abbreviations of datasets.}
\label{fig:label_words}
\vspace{-4mm}
\end{figure}

\section{More Results of Prompt Choices}

\noindent \textbf{Effect of template choices.}
 Figure \ref{fig:templates} gives the experimental results of template choices.
We have two findings:
(i) $\text{T}_2(\cdot)$ performs better overall;
(ii) As mentioned in Section \ref{sec:variants}, continuous templates can bring performance improvement in most cases.

\noindent \textbf{Effect of label words choices.}
The results of label words choices are depicted in Figure \ref{fig:label_words}.
We compare our designed label words with a simple one, i.e., matched for "yes" and mismatched for "no". Both $\text{T}_1(\cdot)$ and $\text{T}_2(\cdot)$ use the continuous versions.
We can observe that the F1-score is significantly dropped using the simple label words.
This demonstrates the effectiveness of our designed label words. In other words, considering the more general relationship between entities is beneficial to the predictions.

\section{Error Analysis}
\begin{figure*}[h]
    \centering
    \includegraphics[width=7in]{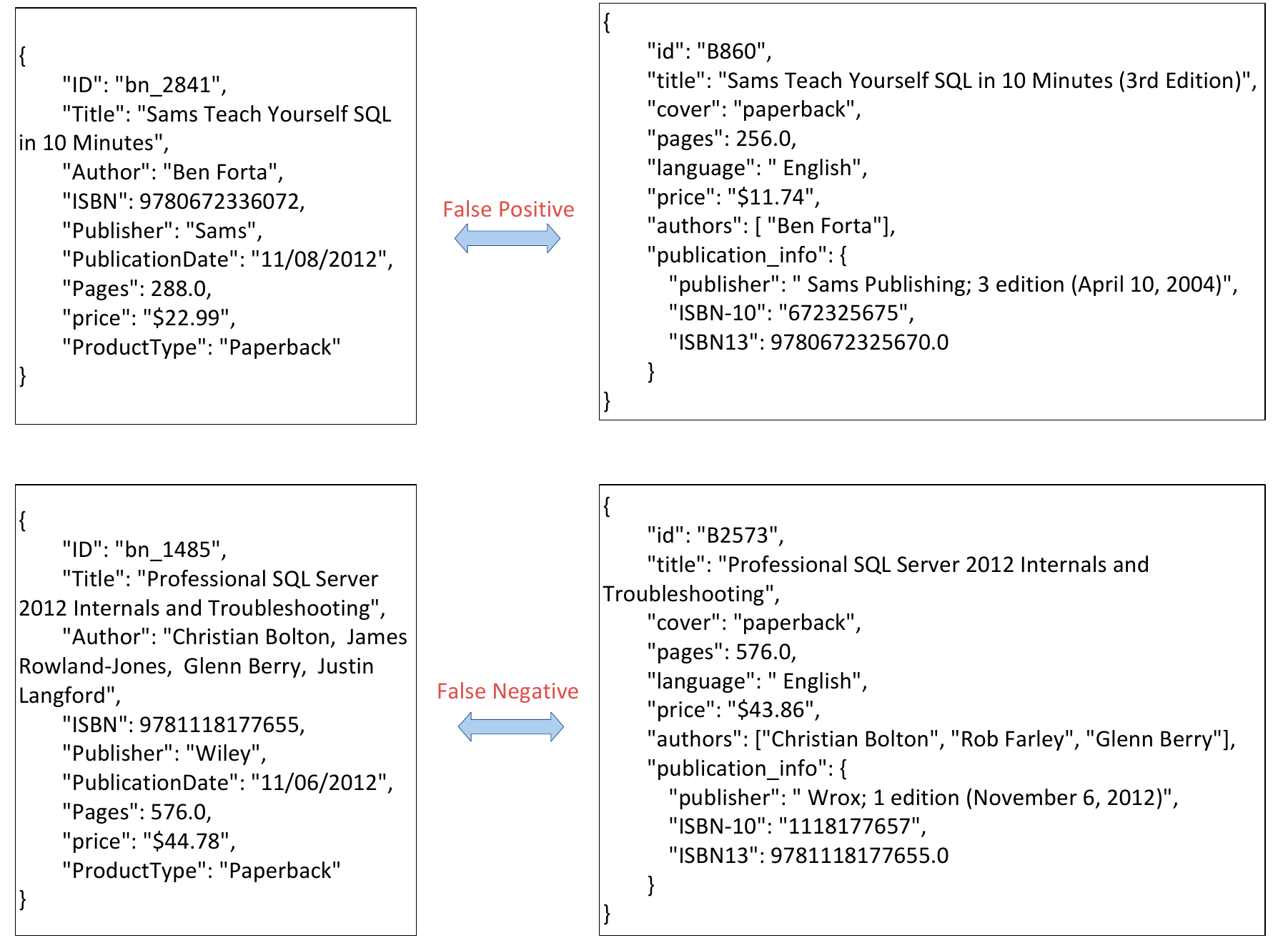}
    \vspace*{-6mm}
    \caption{An error analysis on SEMI-HETER.}
    \vspace*{-1mm}
    \label{fig:error_case}
\end{figure*}

We conduct an error analysis to verify the limitations of our proposed \textsf{PromptEM}.
Figure \ref{fig:error_case} plots the erroneous predictions, including a false positive (FP) and a false negative (FN).
For the FP case, two entities have very similar titles and the same authors. Indeed, we can find that they are different entities based on the publication date and the ISBN.
For the FN case, two entities have the same titles but different authors and publishers. However, we can judge they are the same based on the publication date and the ISBN.
We can find that some digital attributes are vital to making correct predictions.
Nonetheless, \textsf{PromptEM} and other LM-based methods are not good at understanding digits, and hence, they cannot find the importance of those digital attributes.
Therefore, it is necessary to improve the ability of LMs to understand and focus on important digital attributes.

\section{Implementation Details of Baselines}

We implement each baseline method as follows and report its performance under its optimal settings.
\begin{itemize}[topsep=0pt,itemsep=0pt,parsep=0pt,partopsep=0pt,leftmargin=*]
    \item \textsf{DeepMatcher} \cite{mudgal2018deep}: We implement \textsf{DeepMatcher} following the original paper and public code\footnote{\url{https://github.com/anhaidgroup/deepmatcher}}. As \textsf{DeepMatcher} only accepts input with the same schema on two tables, we just make the schemas of both tables have only one attribute, whose value is a sentence consisting of all attribute values. And we use the default setting of the hybrid model.
    \item \textsf{BERT} \cite{devlin2018bert}, \textsf{SentenceBERT} \cite{reimers2019sentence}, \textsf{Ditto} \cite{li2020deep} and \textsf{Rotom} \cite{miao2021rotom}: These methods are based on LMs. We implement these methods according to the original papers, public $\text{code}_{1}$\footnote{\url{https://github.com/megagonlabs/ditto}} and public $\text{code}_{2}$\footnote{\url{https://github.com/megagonlabs/rotom}}. Following \cite{wang2021machamp}, we tune the hyper-parameters by doing a grid search and select the one with the best performance. Specifically, the learning rate is selected from \{$10^{-5}$, $3.0\times10^{-5}$, $5.0\times10^{-5}$\}. The maximum sequence length for LMs is selected from \{128, 256, 384, 512\}. The batch size and epoch are set to 32 and 40, respectively.
    \item \textsf{DADER} \cite{tu2022domain}: We implement \textsf{DADER} following the original paper and public code\footnote{\url{https://github.com/ruc-datalab/DADER}}. For the source dataset, we use all the training samples. For the target dataset, we use the same low-resource training samples as other supervised methods. Specifically, we select the source and target datasets from a similar domain for \textsf{DADER}. We use the same hyper-parameters setting according to the paper. And we use the \textsf{InvGAN+KD} model.
    \item \textsf{TDmatch} \cite{ahmadi2021unsupervised}: We implement \textsf{TDmatch} according to the original paper and public code\footnote{\url{https://github.com/naserahmadi/TDmatch}}. All hyper-parameters are the same as the original paper. To perform in the supervised setting, we extract the embeddings from \textsf{TDmatch}. Then we build a classification layer (i.e., MLP) upon the embeddings, whose name is \textsf{TDmatch*}. Given two entity embeddings $u$ and $v$, we use ($u, v, |u - v|, u * v$) as input for the classifier.
    The batch size is set to 64; the number of epoch is set to 100; and the learning rate is set to $5.0\times10^{-3}$.
\end{itemize}

\begin{table*}[h] \small
\centering
\caption{Results of all the methods under sufficient resource setting.}
\vspace{-3mm}
\label{table:sufficient}
\setlength{\tabcolsep}{0.8mm}{
\begin{tabular}{|c|ccc|ccc|ccc|ccc|ccc|ccc|ccc|ccc|}
\hline
\multirow{2}{*}{\textbf{Methods}} & \multicolumn{3}{c|}{\textbf{REL-HETER}} & \multicolumn{3}{c|}{\textbf{SEMI-HOMO}} & \multicolumn{3}{c|}{\textbf{SEMI-HETER}} & \multicolumn{3}{c|}{\textbf{SEMI-REL}} & \multicolumn{3}{c|}{\textbf{SEMI-TEXT-c}} & \multicolumn{3}{c|}{\textbf{SEMI-TEXT-w}} & \multicolumn{3}{c|}{\textbf{REL-TEXT}} & \multicolumn{3}{c|}{\textbf{GEO-HETER}} \\
                                 & P        & R        & F                 & P        & R        & F                 & P         & R        & F                 & P        & R        & F                & P         & R         & F                 & P         & R         & F                 & P        & R       & F    & P        & R       & F            \\ \hline
\textsf{DeepMatcher}                      & 100    & 87.9     & 93.6              & 89.0     & 83.2     & 86.1              & 35.8      & 24.5     & 29.1              & 50.9     & 64.1     & 56.7             & 76.6      & 31.1      & 44.2              & 80.2      & 29.1      & 42.7              & 78.4     & 40.4    & 53.4             & 88.0 & 85.4 & 86.6 \\
\textsf{BERT}                             & 95.5     & 95.5     & 95.5              & 93.8     & 93.9     & 93.8              & 90.7      & 30.8     & 46.0              & 87.3     & 94.0     & 90.5             & 89.0      & 88.3      & 88.6              & 79.3      & 67.3      & 72.8              & 61.6     & 64.6    & 63.1          & 88.1 & 92.9 & 90.4     \\
\textsf{SentenceBERT}                     & 66.7     & 72.7     & 69.6              & 85.6     & 89.3     & 87.4              & 100     & 53.5     & 69.7              & 47.8     & 77.0     & 59.0             & 85.1      & 75.1      & 79.8              & 52.3      & 48.3      & 50.2              & 37.2     & 29.5    & 32.9           & 87.1 & 92.7 & 89.8    \\ \hline
\textsf{Ditto}                            & 100    & 100    & \textbf{100}             & 94.7     & 91.6     & 93.1              & 84.6      & 48.4     & 61.6              & 95.8     & 86.9     & 91.1             & 82.2      & 81.3      & 81.8              & 63.6      & 66.3      & 64.9              & 65.6     & 60.1    & 62.7          & 92.1 & 88.1 & 90.1     \\
\textsf{DADER}                            & 87.0     & 90.9     & 88.9              & 87.4     & 87.4     & 87.4              & 98.5      & 40.3     & 57.1              & 87.6     & 96.7     & 92.0             & 13.9      & 100     & 24.4              & 11.4      & 100     & 20.5              & 28.3     & 54.1    & 37.2             & 80.7 & 79.6 & 80.2  \\
\textsf{Rotom}                            & 100    & 100    & \textbf{100}             & 94.4     & 95.1     & 94.7              & 45.5      & 32.1     & 37.6              & 97.7     & 91.3     & 94.4             & 92.0      & 89.1      & 90.5              & 80.5      & 68.3      & 73.9              & 69.7     & 62.2    & 65.7           & 90.5 & 87.5 & 90.0    \\ \hline
\textsf{TDmatch}                          & 56.4     & 100    & 72.1              & 93.7     & 42.0     & 58.0              & 97.2      & 88.1     & 92.4              & 97.5     & 85.8     & 91.3             & 69.0      & 10.4      & 18.0              & 42.3      & 14.2      & 21.3              & 80.2     & 47.3    & 59.5             & 72.8 & 73.0 & 72.9  \\ 
\textsf{TDmatch*} & 50.0 & 50.0 & 50.0 & 86.5 & 91.2 & 88.8 & 57.9 & 41.5 & 48.4 & 77.6 & 98.4 & 86.8 & 77.7 & 72.7 & 75.1 & 74.7 & 58.8 & 65.8 & 60.3 & 52.9 & 56.4 & 72.9 & 78.3 & 75.5 \\
\hline
\textsf{PromptEM}                         & 100    & 100    & \textbf{100}    & 96.5     & 96.0     & \textbf{96.3}     & 99.3      & 87.4     & \textbf{93.0}     & 96.7     & 95.6     & \textbf{96.2}    & 93.4      & 90.9      & \textbf{92.1}     & 80.4      & 70.1      & \textbf{74.9}     & 66.7     & 66.0    & \textbf{66.4}    & 92.5 & 92.7 & \textbf{92.6}  \\
\textsf{PromptEM w/o PT}                  & 100    & 100    & \textbf{100}    & 94.3     & 97.1     & 95.7              & 76.2      & 52.2     & 61.9              & 97.2     & 94.0     & 95.6             & 91.8      & 90.3      & 91.0              & 73.4      & 72.0      & 72.7              & 68.9     & 60.4    & 64.4           & 89.2 & 87.0 & 88.1    \\ \hline
\end{tabular}}
\end{table*}

\section{Details of Experimental Datasets}
We use all seven real-world datasets from \textsf{Machamp} \cite{wang2021machamp}. The two tables of each dataset may have different formats (i.e., relational (REL) format, semi-structured (SEMI) format,
or textual (TEXT) format) and different schemas (i.e., homogeneous (HOMO), or heterogeneous (HETER)).

Additionally, we use one real geo-spatial dataset GEO-HETER, which is derived from OSM-FSQ-Pittsburgh \cite{balsebre2022geospatial}. 
For each entity, there are textual attributes and geographical positions.
Following the previous study \cite{wang2021machamp}, the "latitude" and "longitude" of the right table are combined into a single "position" attribute to convert the dataset into a case of heterogeneous schema.

\begin{table*}[t]\small
\centering
\caption{Recent research on task-specific prompt-tuning. Note that, prompt-tuning has not been introduced in the field of data management, including the entity matching task.}
\vspace*{-3.5mm}
\label{table:tasks}
\setlength{\tabcolsep}{3.6mm}{
\begin{tabular}{|c|c|c|}
\hline
\textbf{Paper}  & \textbf{Venue}      & \textbf{Task}                \\ \hline
GPPT: Graph Pre-training and Prompt Tuning to Generalize Graph Neural Networks & SIGKDD 2022 & Graph Neural Network \\ 
\href{https://dl.acm.org/doi/abs/10.1145/3485447.3511921}{Ontology-enhanced Prompt-tuning for Few-shot Learning} &  WWW 2022          & Knowledge Graph Completion \\
\href{https://aclanthology.org/2022.acl-long.174/}{Adversarial Soft Prompt Tuning for Cross-Domain Sentiment Analysis} &  ACL 2022          & Sentiment Analysis \\
\href{https://aclanthology.org/2022.acl-long.80/}{Continual Prompt Tuning for Dialog State Tracking} &  ACL 2022          & Dialogue System \\
\href{https://aclanthology.org/2022.acl-long.424/}{MSP: Multi-Stage Prompting for Making Pre-trained Language Models Better Translators} &   ACL 2022         & Machine Translation \\
\href{https://aclanthology.org/2022.acl-short.17}{The Power of Prompt Tuning for Low-Resource Semantic Parsing} &   ACL 2022         & Semantic Parsing \\
\href{https://dl.acm.org/doi/10.1145/3477495.3532048}{PTAU: Prompt Tuning for Attributing Unanswerable Questions} &    SIGIR 2022        & Question Answering \\
\href{https://dl.acm.org/doi/10.1145/3477495.3531913}{Selective Fairness in Recommendation via Prompts} &  SIGIR 2022          & Recommendation Sysytem \\
\href{https://dl.acm.org/doi/10.1145/3477495.3531746}{Relation Extraction as Open-book Examination: Retrieval-enhanced Prompt Tuning} &  SIGIR 2022          & Relation Extraction \\ 
% \href{https://openaccess.thecvf.com/content/CVPR2022/html/Wang_Learning_To_Prompt_for_Continual_Learning_CVPR_2022_paper.html}{Learning to Prompt for Continual Learning} & CVPR 2022 & Image Classification\\
\href{http://proceedings.mlr.press/v139/radford21a}{Learning Transferable Visual Models From Natural Language Supervision} & ICML 2021 & Image Classification \\\hline
\end{tabular}}
\vspace*{-3mm}
\end{table*}

\section{Summarizing Long Entries}
When the textual data is an extremely long string, it is harder for the LM to understand (e.g., the input to BERT can have at most 512 sub-word tokens). A common practice is to truncate the sequences. Nevertheless, the truncation strategy is not a wise choice because the important information for matching is usually not at the beginning of the sequences. Inspired by \textsf{Ditto} \cite{li2020deep}, we apply a TF-IDF based summarization technique for textual data in our implementation, which retains non-stopword tokens with high TF-IDF scores.

\vspace{2.4mm}
\section{Insights of Method Choices}
We would like to emphasize that some parts of the \textsf{PromptEM} are nice combination of existing methods as each module has our insights.
We detail our insights for the method choices as follows:
(i) \textbf{Why prompt-tuning?}
Existing SOTA EM methods based on LMs achieve considerable performance.
They fine-tune LMs to convert EM to the sequence pair classification task while it is a suboptimal design (Challenge \uppercase\expandafter{\romannumeral 1}, Section 1, Page 2).
To tune pre-trained LMs for GEM better, we introduce the powerful prompt-tuning to bridge the gap between the pre-training and the fine-tuning.
(ii) \textbf{Why uncertainty?} 
The quality of pseudo-labels determines whether self-training can improve performance.
A common strategy is using confidence to select pseudo-labels. However, this strategy has some serious drawbacks, e.g., incorrect predictions can have high confidence scores in poorly calibrated networks (Challenge \uppercase\expandafter{\romannumeral 2}, Section 1, Page 2).
To select high-quality pseudo-labels,
we employ recent advances in Bayesian deep learning to obtain uncertainty estimates of the teacher model for pseudo-labeling and boosting the self-training process.
(iii) \textbf{Why dynamic data pruning?}
Uncertainty-aware self-training is attractive since it can improve performance.
However, the self-training process can be costly.
To be more specific, the labeled data is augmented by the pseudo-labels produced by the teacher model, which may be beneficial to performance but result in a longer training time (Challenge \uppercase\expandafter{\romannumeral 3}, Section 1, Page 2). 
Intuitively, maybe not all training data contribute to boosting the performance of the student model.
Thus, we propose MC-EL2N to quantify the importance of training data, which can be used to prune useless samples dynamically.
Furthermore, prompt-tuning is powerful and attractive, so recently many tasks introduce it to tune the LMs better, as listed in Table \ref{table:tasks}.
\textsf{PromptEM} is the first work in entity matching that introduces powerful prompt-tuning, having the potential to advance the usage of LMs for GEM (including EM).

\end{document}